\documentclass{jfm}
\usepackage{graphicx}
\usepackage{epstopdf, epsfig}
\usepackage{url}

\usepackage{amsmath,amssymb,amstext} 
\usepackage{caption}
\usepackage{subcaption}

\usepackage{textgreek}
\newcommand{\gmu}{\textmugreek}

\usepackage{soul}
\usepackage{color}

\usepackage[figuresright]{rotating}

\usepackage{tikz}
\usetikzlibrary{intersections}
\usetikzlibrary{calc}
\usetikzlibrary{matrix,backgrounds,decorations.pathreplacing,positioning}
\usetikzlibrary{decorations.markings}

\usepackage{tkz-euclide}
\usetkzobj{all}

\tikzstyle{vecArrow} = [thick, decoration={markings,mark=at position
   1 with {\arrow[semithick]{open triangle 60}}},
   double distance=1.4pt, shorten >= 5.5pt,
   preaction = {decorate},
   postaction = {draw,line width=1.4pt, white,shorten >= 4.5pt}]
\tikzstyle{innerWhite} = [semithick, white,line width=1.4pt, shorten >= 4.5pt]

\usepackage[section]{placeins}

\usepackage[utf8]{inputenc}
\usepackage{cleveref}
\crefname{section}{§}{§§}
\Crefname{section}{§}{§§}

\renewcommand{\hl}{}

\shorttitle{Influence of Turbulent Fluctuations on Detonation Propagation} 
\shortauthor{B. McN. Maxwell et al.} 

\title{Influence of Turbulent Fluctuations on Detonation Propagation}

\author
 {
 Brian McN. Maxwell\aff{1}
  \corresp{\email{bmaxwell@uottawa.ca}},
  R. R. Bhattacharjee\aff{1},\\
  S. S. M. Lau-Chapdelaine\aff{1},
  S. A. E. G. Falle\aff{2},
  G. J. Sharpe\aff{3}\\
  \and 
  M. I.  Radulescu\aff{1}
  }

\affiliation
{
\aff{1}
Department of Mechanical Engineering, University of Ottawa,\\ 161 Louis Pasteur, Ottawa, K1N 6N5, Canada
\aff{2}
School of Mathematics, University of Leeds, Leeds, LS2 9JT, UK
\aff{3}
Blue Dog Scientific Ltd., 1 Mariner Court, Wakefield, WF4 3FL, UK
}

\begin{document}

\maketitle




\begin{abstract}

\hl{The present study addresses the reaction zone structure and burning mechanism of unstable detonations.  Experiments investigated mainly two-dimensional methane-oxygen cellular detonations in a thin channel geometry.  The sufficiently high temporal resolution permitted to determine the PDF of the shock distribution, a power-law with an exponent of -3, and the burning rate of unreacted pockets from their edges - through surface turbulent flames with a speed approximately 3-7 times larger than the laminar one at the local conditions. Numerical simulations were performed using a novel Large Eddy Simulation method where the reactions due to both auto-ignition and turbulent transport and treated exactly at the sub-grid scale in a reaction-diffusion formulation.  The model is an extension of Kerstein \& Menon's Linear Eddy Model for Large Eddy Simulation to treat flows with shock waves and rapid gasdynamic transients.   The two-dimensional simulations recovered well the amplification of the laminar flame speed owing to the turbulence generated mainly by the shear layers originating from the triple points and subsequent Richtmyer-Meshkov instability associated with the internal pressure waves.  The simulations clarified how the level of turbulence generated controlled the burning rate of the pockets, the hydrodynamic thickness of the wave, the cellular structure and its distribution.  Three-dimensional simulations were found in general good agreement with the two-dimensional ones, in that the sub-grid scale model captured the ensuing turbulent burning once the scales associated with the cellular dynamics, where turbulent kinetic energy is injected, are well resolved.}

\end{abstract}

\clearpage

\crefname{section}{§}{§§}
\Crefname{section}{§}{§§}

\section{Introduction}
\label{sec.cellular_dynamics}

\begin{figure}
 \begin{center}
  \includegraphics[scale=0.55]{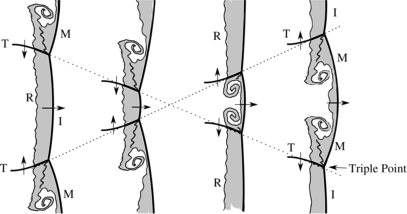}
 \end{center}
 \caption[Sketch showing a triple point collision process.]{Sketch showing a triple point collision process.  Various waves indicated are the incident shocks (I), Mach shocks (M), and transverse shocks (T).  The extent of the turbulent reaction zones are also shown (R).}
\label{fig.detonation_cells_sketch}
\end{figure}

To date, it is well known that multi-dimensional detonations exhibit a characteristic \emph{cellular} pattern on their fronts, as they propagate in a reactive medium.  This cell pattern is associated with wave interactions whose directions are predominantly transverse to the direction of flow.  At each transverse wave collision, whose natural spacing is on the order of 100 ideal, and steady, induction-zone lengths \citep{Fickett1979}, a new cell is formed as triple points are reflected from each other.  This process is illustrated in figure \ref{fig.detonation_cells_sketch}.  The \emph{triple-point} is a key feature of detonation waves.  It is the location where three-principle shock waves meet: the transverse wave, the Mach stem and incident shock.  The two latter are both associated with the front of the wave as indicated in the sketch.  Furthermore, these wave interactions, or cell patterns, are generally classified as either having a \emph{regular} structure or an \emph{irregular} structure. 

\begin{figure}
 \begin{center}
  \includegraphics[scale=0.5625]{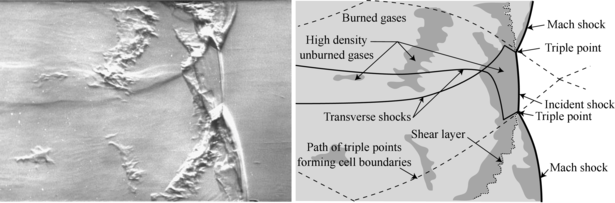}
 \end{center}
\caption[Detonation structure for $\textrm{CH}_4+\textrm{2O}_2$, initially at $\hat{p}_o=3.4$kPa \citep{Radulescu2007b}.]{{Detonation structure for $\textrm{CH}_4+\textrm{2O}_2$, initially at $\hat{p}_o=3.4$kPa.  The top portion shows a Schlieren photograph while the bottom portion includes a corresponding explanatory sketch.} \citep{Radulescu2007b}}
\label{fig.Radulescu2007}
\end{figure}

Regular detonations exhibit very structured-looking fish-scale patterns with consistent, and repeating, cell sizes \citep{Austin2003,Strehlow1968,Pintgen2003}.  In general, regular detonations are associated with fuel mixtures having low-activation energies \citep{Strehlow1968}.  Typical examples are $\textrm{H}_2+\textrm{O}_2$ (pure hydrogen-oxygen) and ${\textrm{C}_2}{\textrm{H}_2}+\textrm{O}_2$ (acetylene-oxygen).  Owing to the relatively low activation energies, regular mixtures typically have {little variation of ignition delays for the reactive gas mixture which passes through the wave front structure}.  For this reason, it is generally believed that the principle \hl{ignition mechanism is by adiabatic shock compression} \citep{Radulescu2005}.  \hl{This was shown to be the case for regular detonations where the structure is well predicted by considering only the ignition delay history of a shocked particle, where transport mechanisms have been neglected} \citep{Edwards1978}.  For \emph{irregular} mixtures, many experiments \citep{Kiyanda2013,Austin2003,Strehlow1968,Radulescu2005,Subbotin1975} have shown that a very different cell pattern exists.  Typically, irregular mixtures correspond to hydrocarbon fuels with oxygen or air.  These mixtures tend to exhibit much more stochastic-looking cell structures.  The cell sizes are much more variable, and sometimes contain what appears to be cells within cells; a smaller cell structure within a much larger, and more prominent, cell structure.  Furthermore, a number of experiments have shown the wave front to be highly turbulent, often giving rise to shocked unburned pockets of reactive fuel in its wake \citep{Subbotin1975,Kiyanda2013,Austin2003,Radulescu2005,Oran1982}.  {An example experiment} \citep{Radulescu2007b} {for an irregular detonation wave, involving methane-oxygen, is shown in figure} \ref{fig.Radulescu2007}.  {In this figure, the turbulent structure is illustrated via Schlieren photography}.  In the same figure, each notable feature is also illustrated and labelled in a supplemental sketch.  These features include the various shock wave dynamics, shear layers, and triple points.  \hl{Of particular interest are the presence of the unburned fuel pockets in the wake, also shown in the figure.  Such pockets contain fine scale corrugations on their edges due to turbulence, where burning occurs.}

{Irregular mixtures, which have higher activation energies, exhibit a much larger variation of shock-induced ignition delays, for unburned fuel passing through the wave structure, compared to regular detonations.  Furthermore, experiments} reveal that the pockets of gas in the wake burn up relatively quickly, orders of magnitude faster than expected from diffusion-less ignition \citep{Radulescu2005,Radulescu2007b,Kiyanda2013}.  As a result, the \hl{Zel'dovich - von Neumann - Doring (ZND) model} \citep{Fickett1979} does not predict well the structure behind the incident and Mach shocks \citep{Radulescu2005}.  This suggests that any unburned pockets in the wake of irregular mixtures burn through turbulent mixing with product gases, and not by shock-compression alone.  Currently, it is not yet understood how the burning rate of these pockets affects the cell pattern observed on the wave front, and is the principal topic investigated in this study.  {One hypothesis proposes that} {as energy is released from the burning of unreacted fuel mixture pockets, pressure pulses can reach and perturb the leading shock, thus influencing the overall structure to generate new cells} \citep{Oran1982}.  Such pressure pulses are able to reach the detonation front since the unreacted pockets lie within the hydrodynamic structure of the detonation, between the leading front and its trailing average sonic surface \citep{Radulescu2007b}.  Furthermore, the triple-point appears to play a major role affecting the dynamics of how these pockets burn out.  Not only is the triple-point a source of {high temperature and pressure} owing to shock-compression from multiple waves, but it is also a source of enhanced turbulent mixing.  The triple point has been shown to give rise to a shear layer in its wake that is susceptible to the Kelvin-Helmholtz (KH) instability and forward jetting \citep{Massa2007,Mach2011,Bhattacharjee2013b}. This shear layer thus acts to enhance mixing between burned product gases with unburned pockets, and therefore acts to increase reaction rates.  \hl{Therefore, to gain further insight on irregular detonation propagation, it is important to model correctly this hybrid ignition regime, where both turbulent mixing and compression ignition are important.}

To date, numerical investigations have been able to show that, with sufficient resolution, mildly irregular regular detonation structures are recovered quite well by solving Euler's equations of fluids motion \citep{Gamezo1999,Sharpe2001,Cael2009}, which do not account for turbulent mixing or molecular diffusion.  More recent numerical investigations, however, have attempted the same approach for modelling highly irregular mixtures, with much higher activation energies, but with limited success \citep{Radulescu2007b}.  Although such attempts have provided some insight into the roles that shock compression or turbulent motions may have on detonation propagation, the solutions obtained were subject to changes in resolution and did not converge to unique solutions \citep{Radulescu2007b}.  Furthermore, such investigations have confirmed that very long induction times exist, \hl{on} the order of one hundred times longer than observed experimentally, and that unburned pockets burn out much more rapidly through numerical diffusion.  Some recent modelling attempts through Direct Numerical Simulation (DNS) of the Navier-Stokes (N-S) equations address this problem by attempting to resolve the full spectrum of scales present, including molecular diffusion effects.  Unfortunately, such investigations \citep{Gamezo2000,Mahmoudi2014} have revealed that practically attainable resolutions, in full-scale {two-dimensional} problems, are insufficient to capture the correct reaction rates of unburned pockets.  Thus, full-scale DNS is currently limited to providing insight only on single-isolated events, such as triple point collisions \citep{Ziegler2011,LauChapdelaine2016}.  {Furthermore, turbulence inherently contains three-dimensional \hl{effects}. It is well known that the dissipation of turbulent motions, the \emph{Kolmogorov energy cascade}, depends intimately on the ability of vortices to stretch in the third dimension.  For this reason, two-dimensional flows tend to experience more backscatter, and hence produce larger-scale fluid motions, compared to realistic three-dimensional flows} \citep{Kraichnan1967,Leith1968,Batchelor1969}.  In order to address these fundamental short comings associated with Euler simulations or DNS, a reasonable compromise between accuracy of solution and resolvability is to employ Large Eddy Simulation (LES).  For LES, the large scale fluid motions, governed by the N-S equations, are solved directly using numerical methods.  The unresolved micro-scale turbulence is then modelled as a supplement to the large scale numerical solutions.  Recent studies have \citep{Mahmoudi2014,Gottiparthi2009} demonstrated that LES can be used to provide insight on the sub-grid turbulent mixing effects that contribute to the highly irregular detonation reaction rate.  Full closure to the reaction rate, however, remains difficult as it is typically obtained by assuming either instantaneous mixing or reaction at the subgrid scale.  {Owing to the difficulties associated with each of these strategies, adequate resolution of the reaction rate of fuel following the wake of a detonation wave remains problematic.}  All of this previous work clearly highlights the need to isolate and resolve turbulent mixing and combustion rates in mixtures prone to irregular structures.

{An alternative LES methodology which shows promise at capturing and resolving mixing rates in detonation propagation problems, is the Linear Eddy Model for Large Eddy Simulation (LEM-LES)} \citep{Menon2011}.  {In the past, LEM-LES has been successfully applied to model weakly compressible turbulent premixed and non-premixed flames} \citep{Menon1996,Chakravarthy2000}.  {The method is therefore adaptable to a range of combustion regimes.  Furthermore, the method has also been successfully applied to model supersonic inert mixing layers} \citep{Sankaran2005}.  Only recently has the methodology been applied to treat, simultaneously, highly compressible and reactive flows that involve very rapid transients in pressure and energy \citep{Maxwell2015,Maxwell2016}; termed Compressible LEM-LES (CLEM-LES).  In the CLEM-LES context, the subgrid is treated as a one-dimensional sample of a diffusion-reaction system within each multi-dimensional LES cell. This reduces the expense of solving a complete multi-dimensional problem through DNS while preserving micro-scale hot spots and their physical effects on ignition. Thus, the model provides high resolution closure for the unresolved chemical reaction terms in the governing, LES-filtered, reactive Navier-Stokes equations. A principle advantage of this methodology is the ability to treat flows where chemical reactions and micro-scale mixing occur on the same scales, without the need to provide compromising assumptions on the reaction rate.

In the current work, experiments and numerical simulations have been conducted in order to identify, qualitatively and quantitatively, the flow instabilities that lead to turbulence in the wake of an irregular detonation wave.  Moreover, the effect that turbulent mixing has on the burning rates of unburned pockets of reactive gas in the wake of such detonations has been studied.  Particular focus has been placed on investigating the impact that such turbulent burning of pockets has on local wave velocities, observed cell patterns, and overall wave structure.  From the physical experiments, useful quantitative statistical properties of the wave front (i.e. wave velocity distributions) have been obtained, which are intended to serve as validation criteria for numerical strategies applied to highly compressible combustion problems involving detonations.  To compliment this experimental work, numerical simulation have been conducted, using the CLEM-LES strategy \citep{Maxwell2016}, in order to examine the effect that turbulent fluctuations have on the resulting flow field evolution, through varying turbulence intensities of the flow.  This was achieved through calibration of a single tuning parameter, as will later be discussed in the paper.  As such, the numerical simulations have been validated to the experiments accordingly.  Further validation has also been made through comparison to similar experiments \citep{Kiyanda2013}, which have also investigated irregular detonation propagation in methane-oxygen.

The paper is organized as follows.  First, the experimental work is presented in \cref{sec.experiments}, which includes the methodology and also qualitative and quantitative results.  \cref{sec.reconstruction} presents the strategy and formulation adopted to reconstruct, numerically, the observed experimental flow field.  \cref{sec.Sim_results} then presents the results obtained through numerical simulation.  Next, \cref{sec.propagation_discussion} offers a discussion on the findings from both experiments and numerical simulations and provides further analysis.  Finally, conclusions are presented in \cref{sec.conclusion}.

\section{Experiments}
\label{sec.experiments}
\subsection{Methodology}

\begin{figure}
  \centering
  \includegraphics[scale=0.5]{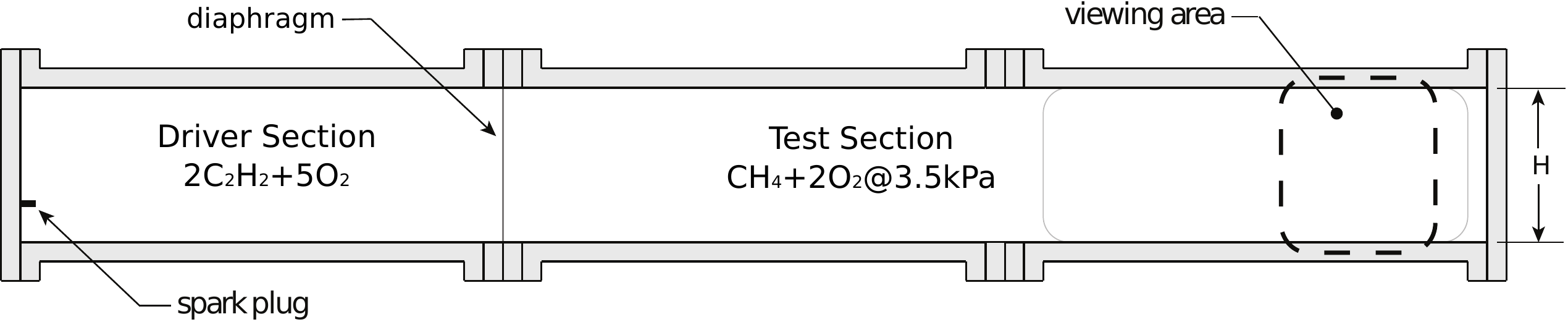}
  \caption{Shock tube set-up for the detonation propagation experiment in $\textrm{CH}_4+\textrm{2O}_2$, initially at $\hat{p}_o=3.5kPa$.}
  \label{fig.LabSetup}
\end{figure}

For the experiments conducted in this study, a shock tube technique is used, as illustrated in figure \ref{fig.LabSetup}.  The shock tube is 3m in total length and has a rectangular cross section whose height is $\hat{H}$=203mm by $\hat{d}$=19mm wide.  The narrowness of its cross section in one direction permits to establish detonations whose cellular structure is essentially two-dimensional, permitting comparison with two-dimensional simulations.  The experimental setup consists of a test section where a methane-oxygen mixture ($\textrm{CH}_4+\textrm{2O}_2$) was filled to the desired pressure of $\hat{p}=3.5\pm$0.1kPa.  Such a low pressure has been chosen to allow the formation of \hl{cells larger than the channel thickness, in order to obtain mainly two-dimensional structures}.  Furthermore, this pressure is consistent with previous investigations \citep{Radulescu2007b,Kiyanda2013}, to provide useful comparison.  This test section was separated from a driver section, by an 80{\gmu}{m} thick plastic diaphragm, which contained acetylene-oxygen (${\textrm{C}_2}{\textrm{H}_2}+\textrm{2.5O}_2$) at $\hat{p}\sim20$ kPa.  A detonation was first initiated in the driver section by a capacitor discharge, delivering $\sim1$kJ in less than 2{\gmu}s \citep{Bhattacharjee2013b}, to a spark plug located at the end wall, as shown in figure \ref{fig.LabSetup}.  This application of a driver section was necessary to ensure that a detonation wave propagates through the test section, which is difficult to initiate directly at such a low pressure.  Prior to conducting the experiment, each chamber was evacuated below 80 Pa before it was filled with the respective gas.  To capture the resulting flow field evolution, a high speed camera (Phantom V1210) was used to take Schlieren photographs, an imaging technique which uses refraction of light in a fluid to capture density gradients \citep{Settles2001}, with a frame rate of 77108fps.  The Schlieren system uses 12 inch field mirrors and a continuous high intensity LED light source.  This permits velocimetry of the various shock speeds in the flow field with $\sim10${\gmu}s time resolution.  Further details on the experiments can be found in \cite{Bhattacharjee2013b}.

\begin{figure}
  \centering
  \includegraphics[scale=0.8]{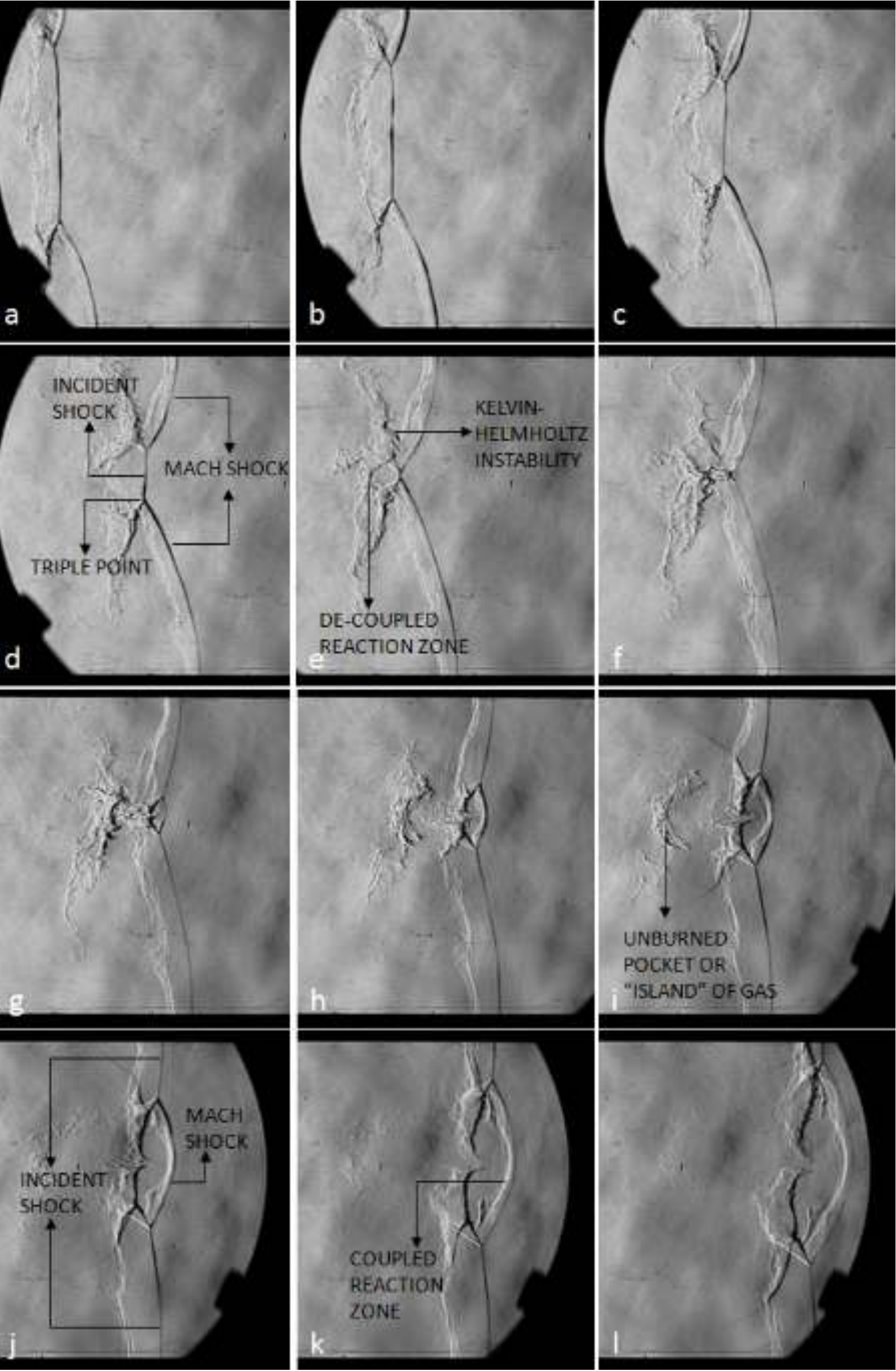}
  \caption[Detonation structure for $\textrm{CH}_4+\textrm{2O}_2$, initially at $\hat{p}_o=3.5kPa$, obtained for two successive frames, 11\gmu s apart \citep{Bhattacharjee2013b}.]{{Detonation flow evolution for $\textrm{CH}_4+\textrm{2O}_2$, initially at $\hat{p}_o=3.5kPa$, obtained for successive frames, 11\gmu s apart} \citep{Bhattacharjee2013b}. {Indicated in the figure are various features of interest; the incident and Mach shocks, triple points, de-coupled reaction zone (pocket of unreacted gas), and shear layers where KH instability is present.}}
  \label{fig:BhattacharjeeOverview}
\end{figure}

\subsection{Qualitative observations}

\label{sec.DetonationPropagation}

In figure \ref{fig:BhattacharjeeOverview}, an example flow evolution of a detonation in $\textrm{CH}_4+\textrm{2O}_2$ at $\hat{p}=3.5\pm$0.1kPa is shown.  This figure shows a sequence of Schlieren images, at 11{\gmu}{s} intervals, where various features are observed.  These features include the incident and Mach shocks, and triple points, previously indicated in figure \ref{fig.detonation_cells_sketch}.  In the sequence of images, presented in figure \ref{fig:BhattacharjeeOverview}, the reaction zone behind the incident shock, labelled in Frame (d), is de-coupled from the shock wave.  This is observed by a thick region between the smooth shock and the textured region where turbulent burning occurs.  This decoupling between the shock and reaction zone can be seen to give rise to a pocket of unburned gas behind the triple point collision process, as observed in frames (e) though (i).  Furthermore, shear layers are observed behind the triple-point trajectories, which give rise to Kelvin-Helmholtz (KH) instability.  These hydrodynamic instabilities are further disrupted and thus enhanced by the passage of transverse shock waves, also originating from the triple point.  Finally, behind the newly formed Mach shocks, following the triple-point collisions, the wave appears overdriven, as observed by a close coupling between the reaction zone and leading shock wave in frame (k).

\begin{figure}
  \centering
  \includegraphics[scale=0.7]{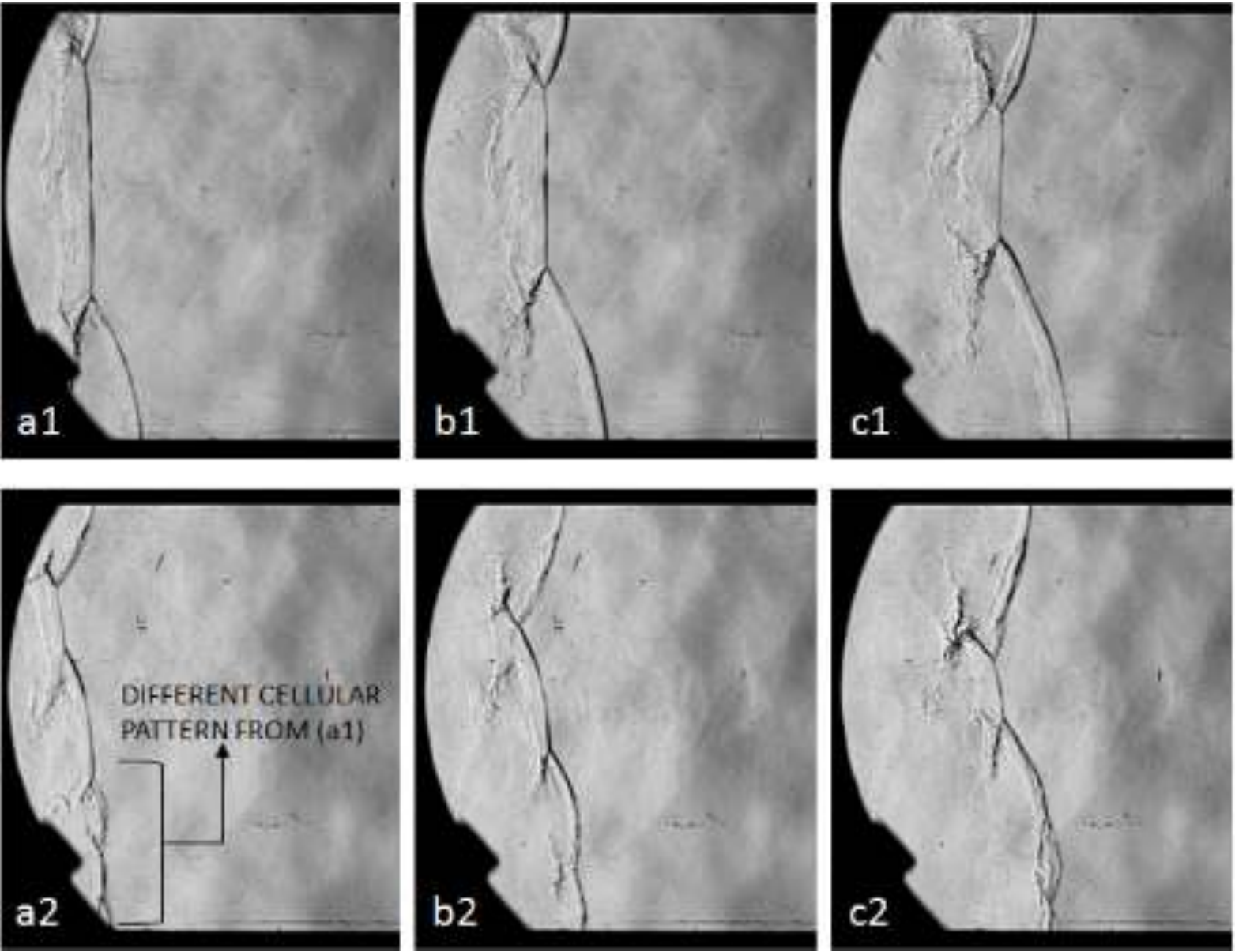}
  \caption[Detonation structure for $\textrm{CH}_4+\textrm{2O}_2$, initially at $\hat{p}_o=3.5kPa$, obtained for two successive frames, 11\gmu s apart \citep{Bhattacharjee2013b}.]{{Another detonation flow evolution for $\textrm{CH}_4+\textrm{2O}_2$, initially at $\hat{p}_o=3.5kPa$, also obtained for successive frames, 11\gmu s apart} \citep{Bhattacharjee2013b}. {In this experiment, a different cell pattern compared to figure \ref{fig:BhattacharjeeOverview} is observed on the wave front.}}
  \label{fig:BhattacharjeeOverview2}
\end{figure}

In figure \ref{fig:BhattacharjeeOverview2}, a repeated experiment at the same conditions reveals a very different cell pattern on the wave front.  Thus the wave structure appears stochastic, reflecting the highly irregular nature of detonations in methane-oxygen.  Thus, it is of particular interest to examine how turbulent mixing, driven through turbulent instabilities and velocity fluctuations, can impact the overall cell structure and propagation characteristics of a detonation wave.

\subsection{Statistical analysis: Velocity of the wave}
\label{sec.expvelocity}

\begin{figure}
  \centering
  \includegraphics[scale=1.2]{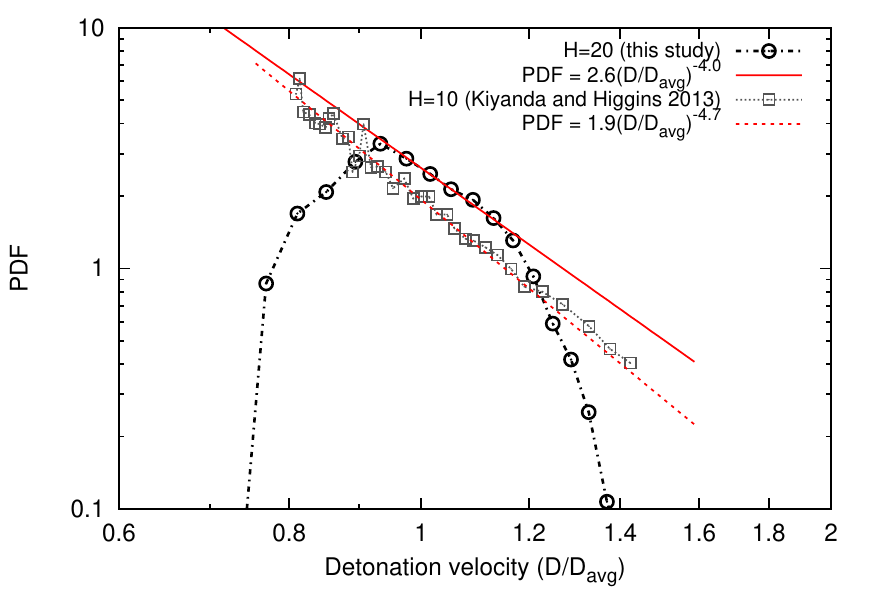}
  \caption[]{PDF of a detonation wave, for $H=20$ (this study), having a certain velocity ($D/D_{avg}$) at any given moment and location. Also shown is a PDF compiled from \cite{Kiyanda2013} for $H=10$.  \hl{Note:  $D_{avg}=5.19$ (1850$\textrm{ms}^{-1}$) for $H=20$ and $D_{avg}=5.53$ (1970$\textrm{ms}^{-1}$) for $H=10$.  The theoretical CJ value is $D_{CJ}=6.30$ (2240$\textrm{ms}^{-1}$).}}
  \label{fig:exp_pdf}
\end{figure}

Four different experiments were conducted at the same conditions.  In order to obtain useful quantitative data, and to gain insight into the observed flow field patterns, velocimetry of the wave was performed on all images obtained from all experiments.  A probability density function (PDF) of wave velocities was constructed and is shown in figure \ref{fig:exp_pdf}.  This figure shows the probability of the wave having a certain velocity at any random time and location.  The wave speeds expected on the detonation front ($D$) are normalized by the averaged propagation speed, which for this study was found to be $D_{avg}=5.19$ \hl{(1850$\textrm{ms}^{-1}$).  It is noted that an average velocity deficit exists compared to the theoretical Chapman-Jouguet (CJ) value of $D_{CJ}=6.30$ (2240$\textrm{ms}^{-1}$), owing to mass divergence from the flow to the boundary layers found on the glass walls of the channel} \citep{Fay1959}.  To measure the wave speed, at any given location on the shock, the distance the shock travels, along its normal vector, to the next intersecting shock location in the subsequent Schlieren image is divided by the time in between images (11{\gmu}s).  A total of 6651 velocity measurements were collected using this procedure.  The PDF was then evaluated at $\delta D=0.2$ intervals.  Also shown in the figure, for comparison, is a PDF obtained from experimental data provided in \cite{Kiyanda2013} for an experiment at the same conditions, but for half of the channel height ($H=10$).  Here, 37 velocity data points are available, \hl{along the channel walls}, to construct the PDF from one entire cell cycle.  \hl{For this experiment, $D_{avg}=5.53$ (1970$\textrm{ms}^{-1}$)}.

Two principal observations are made.  First, for this study ($H=20$), the most probable wave speed favoured is $(D/D_{avg})=0.93\pm0.04$ with a peak PDF value of 3.3.  This implies that the detonation tends to favour a greater chance of having wave speeds below the average propagation speed value, i.e., when $(D/D_{avg})<1$. \hl{This observation was also previously made for irregular detonation wave propagation, through statistical analysis, in several studies} \citep{Radulescu2007b,Shepherd2009,Mevel2015}.   Since the wave spends more time at below CJ-velocities, most of the unburned gas that is shocked by the wave front has a \hl{lower post-shock temperature, and thus a} much longer ignition delay compared to the ZND model.  For this reason, \hl{and owing to very strong unsteady expansion effects} \citep{Lundstrom1969,Austin2003,Radulescu2005,Kiyanda2013}, unburned pockets of gas are able to form in the wake, which eventually burn up through turbulent mixing.  \hl{This burning through turbulent mixing is believed to be a very important mechanism which allows irregular detonations to sustain propagation} \citep{Radulescu2005,Radulescu2007b,Borzou2016}.  The second observation made is that the PDF has a decaying behaviour, where the probability of the wave speed exhibits an approximate power-law dependence on wave speeds above the favoured value.  From the experiments conducted here, the PDF was found to be well predicted by $\textrm{PDF}=2.6(D/D_{avg})^{-4.0}$.  A similar power-law correlation was also observed by \cite{Radulescu2007b} from high resolution Euler simulations, who found a -3 power law dependence of the PDF on detonation velocity.  When compared to existing experimental data \citep{Kiyanda2013}, whose PDF correlation was found to be $\textrm{PDF}=1.9(D/D_{avg})^{-4.7}$, very close agreement is observed, despite propagation through a channel with only half of the height ($H)$.  In \cite{Kiyanda2013}, the most probable wave speed favoured is $(D/D_{avg})=0.814\pm0.005$ with a peak PDF value of 6.15.

\subsection{\hl{Rate of burning of fuel pockets in the wake}}
\label{sec.burningrate}

\hl{For the detonation wave, shown previously in figure} \ref{fig:BhattacharjeeOverview}, \hl{it is of particular interest to estimate the rate at which burning of the pockets of unreacted gas occurs.  Upon isolating a single pocket surface, starting with frame (g) in figure} \ref{fig:BhattacharjeeOverview}, \hl{an approximate method to estimate the turbulent flame speed is to consider the pocket's size, and how long it takes to be consumed. For this method, the pocket's volume ($\hat{V}$) and surface area ($\hat{A}_s$) are estimated from the figure by tracing the pocket shape with a linear spline, multiplied by the channel depth in the third dimension. In this regard, the reader is cautioned that the extent of the structures based on the Schlieren image is sensitive to the experimental setup, for which the sensitivity has not been quantified.  For this particular pocket, the characteristic length of the pocket is estimated as $\hat{L}=\hat{V}/\hat{A}_s=5.30$mm, where $\hat{V}=36,000\textrm{mm}^3$ and $\hat{A}_s=6,800\textrm{mm}^2$.  From this moment, it takes the pocket 4 frames (44\gmu s) to become fully consumed.  Thus, the turbulent flame speed, from experiment, is $\hat{S}_t\approx \hat{L}/\Delta \hat{t}=120\textrm{ms}^{-1}$.  To compare with the local laminar flame speed, post-shock conditions are considered for a shock travelling at 70\% the theoretical CJ speed for the given quiescent mixture.  This particular state of reference (70\% CJ) has been chosen since wave velocity deficits of 20-30\% have been found in the current experiment to exist near the end of the cell cycle during propagation. Furthermore, this state is representative of the pockets of unreacted gas, which form in regions where the shock strength is weaker.  The velocity deficits, near the end of the cell cycle, give rise to a significant increase in ignition delay by several orders of magnitude} \citep{Radulescu2007b}, \hl{thus allowing the pocket to decouple from the wave front. Then, by considering realistic chemistry using Cantera libraries} \citep{Goodwin2013} \hl{and the GRI-3.0 detailed kinetic mechanism} \citep{Smith2013}, \hl{the state properties of the unburned pocket, and the laminar flame speed are obtained prior to auto-ignition of the pocket.  In this case, $\hat{S}_{L,70\%CJ}=16.4\textrm{ms}^{-1}$.  Thus, the turbulent flame speed is found to be approximately seven times larger than the laminar flame speed, i.e., $\hat{S}_t/\hat{S}_L\approx7.3$.}

\section{Numerical Reconstruction of the Flow Field}
\label{sec.reconstruction}
\subsection{Overview}

\hl{In order to reconstruct the flow field evolution observed experimentally, the Linear Eddy Model for Large Eddy Simulation} \citep{Menon2011} \hl{has been applied.  In this approach, solutions to the governing Navier-Stokes (N-S) equations were obtained at two different scales; the \emph{supergrid} and the \emph{subgrid} scales.  On the large scales, the governing LES equations are obtained in the usual manner by filtering the N-S equations to some attainable scale.  Subgrid contributions are then accounted for through the appropriate closure model.  For LEM-LES, the closure model is a self-contained one-dimensional diffusion-reaction system within each LES cell.} This strategy has proven to be effective for both non-premixed \citep{McMurtry1992,Menon1994,Menon1996} and premixed \citep{Menon1992,Menon1993,Smith1996,Smith1997,Calhoon1995,Calhoon1996,Chakravarthy2000,Porumbel2006} combustion applications.  The LES formulation adopted here largely follows previous LEM-LES implementations \citep{Smith1997,Menon2011,Sankaran2003} with a few differences in formulation and implementation.  The most significant and novel contribution to the advancement of the LEM-LES strategy is the treatment of pressure on the subgrid, and its influence on reaction rates, which allows for adequate closure of the reaction rate in the presence of shocks and strong expansions that evolve on the supergrid.

A significant realization of this methodology is that information regarding local hot spots and contact surfaces on the molecular level are accounted for and resolved. This subgrid modelling strategy thus provides high resolution closure to the filtered reaction rate on the large scales.  Furthermore, to simulate the effect of turbulent mixing at the subgrid scale, the one-dimensional flow fields are randomly ``stirred" by linear eddies according to a Probability Density Function (PDF) \citep{Kerstein1991b}.  The effect of the resulting compressive strain on the flow field, due to the presence of the random subgrid eddies, is to reproduce the turbulent diffusivity on the small scales, based on local properties of the flow on the large scales.  The principal advantage of this modelling strategy is that restrictive assumptions, such as infinitely fast mixing or chemical reactions, are not required. 

Finally, Adaptive-Mesh-Refinement (AMR) is applied to the supergrid, providing increased efficiency in obtaining solutions by computing high resolution solutions only in the regions of interest.  In this section, the strategy formulation is first summarized for both the supergrid and subgrid scales, respectively in \cref{sec.les_eqns_super} and \cref{sec.les_eqns_subgrid}.  Then, the simulation setup, including initial and boundary conditions, and also the numerical parameters are presented in \cref{sec.Detonation_domainparams}.  Finally, results of the simulation are provided in \cref{sec.Sim_results}.  Specific details of the CLEM-LES strategy, including procedural algorithms, numerical methods, and limitations are given in \cite{Maxwell2016}.

\subsection{The filtered LES equations}
\label{sec.les_eqns_super}

For flows which are highly transient, turbulent, compressible, and involve rapid combustion chemistry, the gasdynamic evolution is governed by the compressible N-S equations.  In order to address the difficulty of resolving the full spectrum of length scales resulting from the presence of large flow velocities with high Mach numbers ($M_a$) and Reynolds numbers ($\Rey$), the unresolved scales of the governing equations are filtered and modelled through the LES approach. In this respect, rapid transients and fluid motions are captured on the large scales, while the small scale contributions are modelled through source terms. The LES-filtered conservation equations for mass, momentum, and energy (sensible + kinetic) of a calorically perfect reactive fluid system are given below in equations \eqref{eqn.LESmass2}-\eqref{eqn.LESenergy2}, respectively.  Here, a linear-eddy-viscosity model \citep{Pope2000} and gradient driven turbulent heat diffusion hypothesis \citep{Poinsot2005,Combest2011} have been applied to describe the inert effect of velocity fluctuations on the flow-field in terms of a turbulent kinematic viscosity, $\nu_t$.  The set of equations are further supplemented by a one-equation Localized Kinetic energy Model (LKM) \citep{Schumann1975,Chakravarthy2001} to describe the diffusion, advection, production, and dissipation of the subgrid kinetic energy (${k}^{sgs}$) associated with subgrid velocity fluctuations in the flow, see equation \eqref{eqn.LESke}.  Finally, the equations of state are given by equations \eqref{eqn.EOS}.  It should be noted that the equations are given in non-dimensional form where the various gas properties are normalized by the reference quiescent state. Favre-average filtering is achieved by letting $\tilde{f} = \overline{\rho f} / \bar{\rho}$, where $f$ represents one of the many state variables.  Here $\rho$, $p$, $e$, $T$, and $\boldsymbol{u}$ refer to density, pressure, specific sensible + kinetic energy, temperature, and velocity vector, respectively.  Other usual properties to note are the heat release, $Q$, the ratio of specific heats, $\gamma$, the kinematic viscosity, $\nu$, the resolved shear stress tensor, $\bar{\bar{\boldsymbol{\tau}}}$, and the Prandtl number, $\Pran$.  
\begin{equation}
   {\frac{\partial \bar{\rho}}{\partial {t}}} + {\boldsymbol{\nabla} \cdot (\bar{\rho} \tilde{{\boldsymbol{u}}})} = 0
   \label{eqn.LESmass2}
\end{equation}%
\begin{equation}
   {\frac{\partial (\bar{\rho} \tilde{\boldsymbol{u}})}{\partial {t}}} + {\boldsymbol{\nabla} \cdot (\bar{\rho} \tilde{\boldsymbol{u}} \tilde{{\boldsymbol{u}}})} + {\nabla \bar{p}} - {\boldsymbol{\nabla} \cdot {\bar{\rho}(\nu + \nu_{t})} \biggl( \nabla \tilde{\boldsymbol{u}} + (\nabla \tilde{\boldsymbol{u}})^T - \frac{2}{3}( \boldsymbol{\nabla} \cdot \tilde{{\boldsymbol{u}}} ) \mathsfbi{I} \biggr)} = 0
   \label{eqn.LESmomentum2}
\end{equation}%
\begin{equation}
   {\frac{\partial (\bar{\rho} \tilde{e})}{\partial {t}}} + {\boldsymbol{\nabla} \cdot \biggl((\bar{\rho}\tilde{e}+\bar{p})\tilde{{\boldsymbol{u}}} - \tilde{{\boldsymbol{u}}} \cdot \bar{\bar{\boldsymbol{\tau}}}\biggr)} - {\biggl(\frac{\gamma}{\gamma - 1} \biggr) \boldsymbol{\nabla} \cdot \biggl( \bar{\rho} (\frac{\nu}{\Pran} + \frac{\nu_t}{\Pran_t}) \nabla \tilde{T} \biggr)} = - {Q \overline{\dot{\omega}}}
   \label{eqn.LESenergy2}
\end{equation}%
\begin{equation}
   {\frac{\partial (\bar{\rho} {{k}^{sgs}})}{\partial {t}}} + {\boldsymbol{\nabla} \cdot (\bar{\rho} \tilde{{\boldsymbol{u}}} {{k}^{sgs}})} - {\boldsymbol{\nabla} \cdot \biggl({\frac{\bar{\rho} \nu_{t}}{\Pran_t} \nabla {{k}^{sgs}}} \biggr)} = {\dot{P}} - {\bar{\rho} \epsilon}
\label{eqn.LESke}
\end{equation}%
\begin{equation}
  \tilde{e} = {\frac{{\bar{p}}/{\bar{\rho}}}{(\gamma - 1)}} + {\frac{1}{2} \tilde{\boldsymbol{u}} \cdot \tilde{\boldsymbol{u}}} + { {{k}^{sgs}}} \;\;\;\;\;\;\;\; \mbox{and} \;\;\;\;\;\;\;\;  \bar{\rho} \tilde{T} = \bar{p}
\label{eqn.EOS}
\end{equation}%
Above, the various state variables have been normalized such that
\begin{equation}
\rho = \frac{\hat{\rho}}{\hat{\rho_o}}, \;\;\;\; \boldsymbol{u} = \frac{\hat{\boldsymbol{u}}}{\hat{c_o}},  \;\;\;\; p = \frac{\hat{p}}{\hat{\rho_o} {\hat{c_o}}^2} = \frac{\hat{p}}{\gamma \hat{p_o}}, \;\;\;\; T = \frac{\hat{T}}{\gamma \hat{T_o}}, \;\;\;\; x = \frac{\hat{x}}{\hat{\Delta}_{1/2}}, \;\;\;\; t = \frac{\hat{t}}{\hat{\Delta}_{1/2} / \hat{c_o}}
\label{eqn.nonDim1}
\end{equation}
where the subscript `o' refers to the reference state, the hat superscript refers to a dimensional quantity, $\mathsfbi{I}$ is the identity matrix, $c$ is the speed of sound, and $\hat{\Delta}_{1/2}$ is a reference length scale.  This reference length scale is taken as the theoretical half-reaction length associated with the steady ZND solution for a detonation wave propagating in the quiescent reference fluid.  In the equation set above, the subgrid kinetic energy, $k^{sgs}$, is produced at the same rate from which large-scale turbulent motions are dissipated, on the supergrid, through the turbulent kinematic viscosity.  Hence, the rate of production of subgrid kinetic energy is given by
\begin{equation}
\dot{P} = \bar{\rho}\nu_{t} \biggl( \nabla \tilde{\boldsymbol{u}} + (\nabla \tilde{\boldsymbol{u}})^T - \frac{2}{3}( \boldsymbol{\nabla} \cdot \tilde{{\boldsymbol{u}}} ) \mathsfbi{I} \biggr) \cdot (\nabla \tilde{\boldsymbol{u}})
\label{eqn.turbulentViscosity}
\end{equation}%
and the dissipation rate is modelled as 
\begin{equation}
\epsilon = {\pi\biggl(\frac{2{k}^{sgs}}{3C_\kappa}\biggr)^{3/2}}/{\bar{\Delta}}
\label{eqn.turbulentDissipation}
\end{equation}%
Finally, a Smagorinsky-type model is applied to link the turbulent kinematic viscosity to the unresolved velocity fluctuations \citep{Pope2000}, and thus the subgrid kinetic energy, through 
\begin{equation}
\nu_{t} = \frac{1}{\pi}\biggl(\frac{2}{3C_\kappa}\biggr)^{3/2}\sqrt{k^{sgs}}\bar{\Delta}
\label{eqn.turbulentViscosity}
\end{equation}%
Here, $C_{\kappa}$ is the \emph{Kolmogorov constant}, a model parameter which requires calibration.  Typically, $C_\kappa$ is estimated from experiments to be $\sim 1.5$, however published values range anywhere from 1.2 to 4 \citep{Chasnov1991}.  Also found in equations \eqref{eqn.turbulentDissipation} and \eqref{eqn.turbulentViscosity} is  the LES filter size, $\bar{\Delta}$.  For simplicity, it is assumed that $\bar{\Delta}=b$, where $b$ is the minimum grid spacing.  It is noted, however, that this assumption may introduce errors at fine-coarse cell interfaces when coupled with AMR \citep{Pope2004,Vanella2008}.  Since the bulk of the subgrid kinetic energy is generated and dissipated in regions containing shock waves, which are refined to the highest level, it is believed that such fine-coarse interface errors will not affect the solution outcome.  Finally, to close the conservation of energy equation \eqref{eqn.LESenergy2}, the heat release term, $\overline{\dot{\omega}}$, requires closure.  This is done through the application of the CLEM subgrid within each LES cell.

\subsection{The CLEM subgrid model}
\label{sec.les_eqns_subgrid}
\hl{For the CLEM subgrid modelling strategy, the small scale mixing and chemical reactions are solved separately from the large scale pressure evolution. The present implementation of the LEM strategy, introduced by} \cite{Kerstein1988,Kerstein1989,Kerstein1990,Kerstein1992a,Kerstein1991a,Kerstein1991b,Kerstein1992b}, \hl{differs from previous subgrid formulations for LES} \citep{McMurtry1992,Menon1993,Menon1996,Calhoon1996,Mathey1997,Sankaran2003,Porumbel2006,Menon2011} \hl{by ensuring appropriate coupling of the pressure and energy fields on both scales.}  More specifically, the supergrid simulation, equations \eqref{eqn.LESmass2} through \eqref{eqn.LESke}, provides the local rates of change of pressure to the subgrid (i.e., the pressure evolution), while the subgrid model is then applied to simulate the small scale molecular mixing and chemical reactions, which thus provides $\overline{\dot{\omega}}$ to the supergrid as a source term.  Thus, the sole purpose of the subgrid simulation is to provide closure to $\overline{\dot{\omega}}$ in equation \eqref{eqn.LESenergy2}.

Here, the subgrid model is a one-dimensional representation of the flow field within each {supergrid} cell whose orientation is aligned in the direction of local flow.  Furthermore, the subgrid simulation is only applied to the finest cells on the AMR-enabled supergrid. {Thus, each supergrid cell on the finest grid level, which requires closure, contains one subgrid domain}.  To formally derive the subgrid model, presented below, the low-Mach number approximation \citep{Paolucci1982} is applied to the governing N-S equations.  In this approximation, pressure gradients are locally neglected within the small scales of each {supergrid} cell.   This effectively assumes that pressure waves travel much faster than the physical expansion or contraction of the local fluid relative to its convective motion \citep{Maxwell2015}.

Furthermore, the chemistry has been simplified by assuming that reactants form products according to a single-step global reaction with no intermediate reactions:  $Reactants\rightarrow Products$ directly and irreversibly.  In reality, the reaction rate (${\dot{\omega}}$) is governed by chemistry involving multiple reactions between multiple species.  This requires significant computational overhead due to the requirement of accounting for {hundreds, or thousands, of equations and reactions in reactive flows}.  Thus, for the subgrid system, the resulting conservation of enthalpy and reactant mass, along particle paths, are expressed in non-dimensional form as:
\begin{equation}
   \underbrace{\rho \frac{\textrm{D}T}{\textrm{D}t}}_{\substack{\text{rate of change}\\\text{of enthalpy along}\\\text{a particle path}}} - \underbrace{\biggl(\frac{\gamma - 1}{\gamma} \biggr)\dot{p}}_{\substack{\text{rate of change of}\\\text{energy due to local}\\\text{pressure changes}}} - \underbrace{\rho \frac{\partial}{\partial m}\biggl(\rho \alpha \frac{\partial T}{\partial m} \biggr)}_{\substack{\text{heat diffusion}\\\text{to neighbouring}\\\text{fluid elements}}} = - \underbrace{\biggl(\frac{\gamma - 1}{\gamma} \biggr) Q \dot{\omega}}_{\text{heat release}} + \dot{F_T}
\label{eqn.lemEnergy}
\end{equation}%
\begin{equation}
   \underbrace{\rho \frac{\textrm{D}Y}{\textrm{D}t}}_{\substack{\text{rate of change of}\\\text{reactant density along}\\\text{a particle path}}} - \underbrace{\rho \frac{\partial}{\partial m}\biggl(\rho \frac{\alpha}{Le} \frac{\partial Y}{\partial m} \biggr)}_{\substack{\text{reactant diffusion}\\\text{to neighbouring}\\\text{fluid elements}}} = \dot{\omega} + \dot{F_Y}
\label{eqn.lemReactant}
\end{equation}%
\hl{where $Y$ is defined as the reactant mass fraction.}  In this formulation, the material derivative has been applied, where $\frac{\textrm{D}\phi}{\textrm{D}t}=\frac{\partial\phi}{\partial t}+u\frac{\partial \phi}{\partial x}$.  Also, $m$ is a one-dimensional mass weighted Lagrangian coordinate whose transformation to Cartesian spatial coordinates is given by
\begin{equation}
   m(x,t) = \int_{x_o}^x{\rho(x,t)dx}
\label{eqn.transform}
\end{equation}%
Then, for the one-step global reaction mechanism, which considers a single premixed reactant, with mass fraction $Y$, products are formed according to the single step Arrhenius expression \citep{Williams1985}:
\begin{equation}
    {\dot{\omega}} = -{\rho} {A} Y \textrm{e}^{(-{E_a}/{T})}
\label{eqn.oneStep}
\end{equation}%
where ${E_a}$ is the \hl{non-dimensional} activation energy of the reactant mixture.  Here, the activation energy ($\hat{E_a}$), heat release ($\hat{Q}$), and pre-exponential factor ($\hat{A}$) are normalized \hl{according to}
\begin{equation}
E_a = \frac{\hat{E_a}}{{\hat{c_o}}^2}, \;\;\;\;\;\;\;\; Q = \frac{\hat{Q}}{{\hat{c_o}}^2}, \;\;\;\;\;\;\;\; A = \frac{\hat{A}}{\hat{c_o} / \hat{\Delta}_{1/2}}
\label{eqn.nonDim2}
\end{equation}
Furthermore, it is useful to define the following transport relationships in order to relate viscosity to heat and mass diffusion in terms of Reynolds ($\Rey$), Prandtl ($\Pran$), Schmidt ($Sc$), and Lewis numbers ($Le$):
\begin{equation}
\mu = \rho\nu = \frac{1}{\Rey} = \frac{\hat{\mu}}{\hat{{\rho}_o} \hat{c_o} \hat{\Delta}_{1/2}}, \;\;\;\;\;\; \alpha = \frac{\mu}{\Pran} = \frac{\hat{k} / \hat{c_p}}{\hat{{\rho}_o} \hat{c_o} \hat{\Delta}_{1/2}}, \;\;\;\;\;\; Le = \frac{Sc}{\Pran} = \frac{\hat{k} / \hat{c_p}}{\hat{\rho} \hat{D}}
\label{eqn.nonDim3}
\end{equation}
The source terms, $\dot{F_T}$ and $\dot{F_Y}$, in equations \eqref{eqn.lemEnergy} and \eqref{eqn.lemReactant}, \hl{do not take on any specific values, but rather} account for the effect of turbulence on the subgrid in the form of random ``stirring" events \citep{Kerstein1991b}.   \hl{These ``stirring" events are implemented as a series of random instantaneous re-mapping procedures on the subgrid.  The remapping procedure is designed to simulate the effect that a multi-dimensional eddy would have on a one-dimensional sample of the flow field.  In order to achieve this, a \emph{triplet map} {is} implemented, as detailed by} \cite{Kerstein1991b}.  \hl{Functionally, the application of these ``stirring" events depend on the local turbulent diffusivity (or viscosity), which in turn depends on local velocity fluctuations through the subgrid kinetic energy, $k^{sgs}$.} Finally, the source term involving $\dot{p}$, accounts for enthalpy changes which arise from local temporal changes in pressure, which is obtained from the supergrid simulation.

\subsection{\hl{Numerical implementation}}
\label{sec.implementation}

\hl{In order to solve the system of equations} \eqref{eqn.LESmass2} through \eqref{eqn.LESke}, \hl{a numerical framework developed by Mantis Numerics Ltd. is employed.  The compressible flow solver uses a second order accurate \emph{exact} Godunov solver} \citep{Falle1991,Richtmyer1967}, \hl{which features a symmetric monotonized central flux limiter} \citep{vanLeer1977}, \hl{to treat the convection terms. The diffusive terms are handled explicitly in time using the Forward Euler method, and spatially discretized using second order accurate central differences} \citep{Tannehill1997}.  \hl{Structured Cartesian grids are applied in order to take advantage of Adaptive Mesh Refinement (AMR)} \citep{Falle1993} \hl{for increased efficiency.  It should be noted, however, that the application of LEM-LES may not necessarily be limited to structured Cartesian grids} \citep{Cannon2001}. \hl{In this work, AMR is implemented only on the supergrid, equations} \eqref{eqn.LESmass2} - \eqref{eqn.LESke}\hl{. Specifically, the supergrid is refined, on a per cell-basis, in regions where the density or reactant mass ($\bar{\rho}$ or $\bar{\rho} \tilde{Y}$) changes by more than 0.1\% locally between existing grid levels.  Furthermore, when a cell is flagged as `bad', or needing refinement, this badness is diffused by approximately 5-10 cells in each direction on the current grid level.  For the purpose of refinement, the Favre-averaged reactant mass, $\bar{\rho} \tilde{Y}$, is obtained from information stored entirely on the subgrid.  The supergrid is also refined in the presence of shocked and unburned reactant (i.e., $\bar{\rho} \tilde{Y} > 0$ \& $\bar{\rho} > 1.1$). Furthermore, the subgrid model is only active on the finest cells of the supergrid system, and not the coarsest and intermediate grid levels. AMR is not applied to the subgrid domains.}  \hl{For newly refined supergrid cells, and thus newly created LEM subgrid domains, the initial values of $\rho$, $T$, and $Y$ take on those of the supergrid.}

\hl{To solve the subgrid system of equations} \eqref{eqn.lemEnergy} and \eqref{eqn.lemReactant}, \hl{operator splitting} \citep{Leveque2002} \hl{is applied to treat the various terms.  First, the enthalpy contribution due to pressure changes is applied to each LEM element through the source term $\dot{p}$, as detailed in} \cite{Maxwell2016}.  \hl{Then, the diffusion terms are solved in the same manner as the supergrid diffusion terms: explicitly in time using the Forward Euler method, and spatially discretized using second order accurate central differences} \citep{Tannehill1997}.  \hl{This is done for successive and sufficiently small subgrid time-steps ($\Delta t_\textrm{diff}$) in order to satisfy either the CFL stability criteria associated with the explicit diffusion operation, or the time at which the stirring event occurs, whichever is smallest. Then, the reaction terms are solved, implicitly, for the same time step, $\Delta t_\textrm{diff}$, using the Backward Euler method} \citep{Tannehill1997}. 

\hl{Procedurally, the supergrid system of equations,} \eqref{eqn.LESmass2} through \eqref{eqn.LESke}, \hl{is solved first for one global time step ($\Delta{t}$) without chemical reaction by letting $\overline{\dot{\omega}}=0$.  This provides $\dot{p}$ to the subgrid system, equations} \eqref{eqn.lemEnergy} and \eqref{eqn.lemReactant}, \hl{which is then simulated in each refined supergrid cell for the same $\Delta{t}$.  Zero-gradient boundary conditions are assumed for the LEM domains during this process. This provides the exact $\overline{\dot{\omega}}$ contribution, which is then accounted for in equation} \eqref{eqn.LESenergy2}. \hl{Finally, once the LEM domains evolve across each supergrid time step ($\Delta{t}$), large scale advection of LEM elements is handled by transferring elements to or from neighbouring cells through a splicing procedure detailed in} \cite{Sankaran2003}. \hl{This is easily implemented and satisfies the basic mass conservation requirement since the mass flux ($\bar{\rho}\tilde{\boldsymbol{u}}$) across each LES cell face is known. More specific details on the procedure, methodology, algorithms, and model derivation are found in} \cite{Maxwell2016}.

\subsection{Numerical domain and model parameters}
\label{sec.Detonation_domainparams}

\begin{figure}
  \centering
  \includegraphics[scale=0.6]{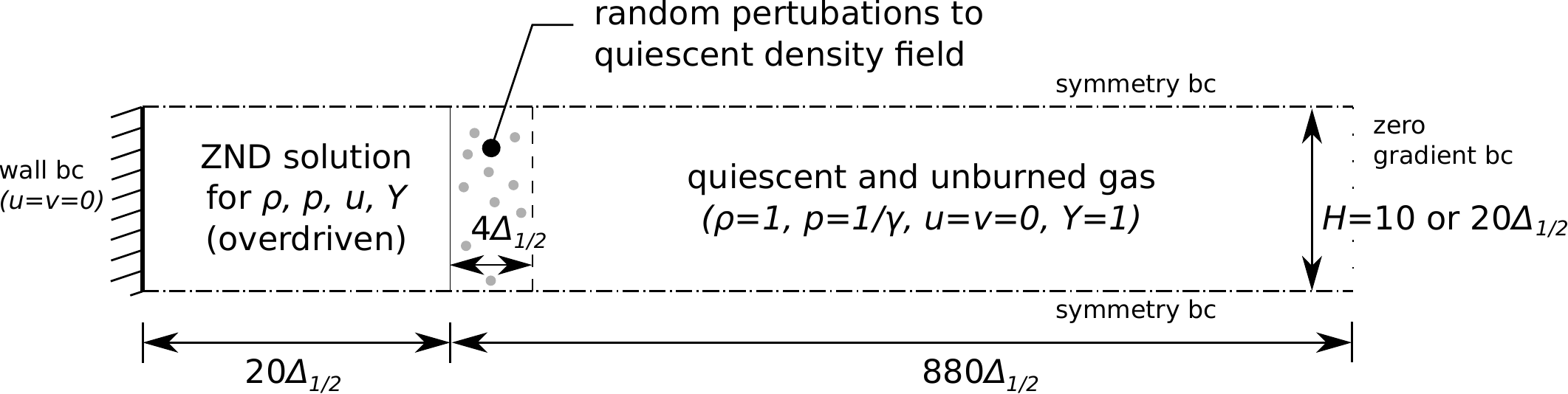}
  \caption[Initial and boundary conditions for the two-dimensional planar detonation propagation experiment.]{Initial and boundary conditions for the two-dimensional planar detonation propagation experiment (Not to scale).  Note: $\hat{\Delta}_{1/2}=9.68$mm.}
  \label{fig:ch4det_icbc}
\end{figure}

A schematic showing the setup for the numerical \hl{domain} is shown in figure \ref{fig:ch4det_icbc}.  In all \hl{simulations}, the total domain length is 900 half reaction lengths ($\hat{\Delta}_{1/2}$) long, which is sufficient to allow adequate sampling and analysis of the detonation structure beyond the time it takes for the CJ-speed to be recovered.  A domain height of 20$\hat{\Delta}_{1/2}$ corresponds to the experiments in \cref{sec.experiments}, while a height of 10$\hat{\Delta}_{1/2}$ corresponds to the experiments of \cite{Kiyanda2013}.   For the wall boundary, shown in figure \ref{fig:ch4det_icbc}, $u=v=k^{sgs}=0$ and the normal gradients of all remaining variables are zero.  For the symmetry boundary condition, only the normal velocity component is taken as zero.  Thus $v=0$, while the normal gradients of all other variables are also zero.  Finally, the zero gradient boundary condition assumes that all normal gradients for all variables are zero.  To initiate the planar detonation wave, an initially over-driven ZND-profile is initialized in the first 20$\hat{\Delta}_{1/2}$ lengths of the computational domain.  The initial ZND-profile corresponds to an over-driven detonation by 10\% or 20\% in order to overcome start-up errors associated with the initially sharp discontinuity at the shock.  A section ahead of the ZND-profile, 4$\hat{\Delta}_{1/2}$ long, contains random perturbations to the quiescent density field.  This allows for the two-dimensional cellular structure to develop further down the channel.  \hl{Finally, the resulting expansion wave which originates from the wall boundary acts to decelerate the wave from its overdriven state to the theoretical CJ speed} \citep{Fickett1979}. Formally, the initial conditions for $\rho$, $u$, $p$, and $Y$ are given according to:
\begin{equation}
\rho(x,y,0) = \left\{ 
  \begin{array}{l l}
    \textrm{ZND solution} & \quad \textrm{if $x<0$}\\
    1.25-0.5n & \quad \textrm{if $0\leq x\leq 4$}\\
    1 & \quad \textrm{otherwise}
  \end{array} \right.
\nonumber
\end{equation}%
\begin{equation}
u(x,y,0) = \left\{ 
  \begin{array}{l l}
    \textrm{ZND solution} & \quad \textrm{if $x<0$}\\
    0 & \quad \textrm{otherwise}
  \end{array} \right.
\end{equation}%
\begin{equation}
v(x,y,0) = 0
\nonumber
\end{equation}%
\begin{equation}
p(x,y,0) = \left\{ 
  \begin{array}{l l}
    \textrm{ZND solution} & \quad \textrm{if $x<0$}\\
    1/\gamma  & \quad \textrm{otherwise}
  \end{array} \right.
\nonumber
\end{equation}%
\begin{equation}
Y(x,0) = \left\{ 
  \begin{array}{l l}
    \textrm{ZND solution} & \quad \textrm{if $x<0$}\\
    1 & \quad \textrm{otherwise}
  \end{array} \right.
\nonumber
\end{equation}%
where $n$ is a random real number from 0 to 1.  For the turbulent kinetic energy, $k^{sgs}(x,y,0) = 0$ everywhere, and is therefore self-generating throughout the simulation.

Next, in order to determine the required model parameters that mimic the experimental premixed methane-oxygen mixtures, from \cref{sec.experiments} and \cite{Kiyanda2013}, the various dimensional and non-dimensional properties were obtained using the procedure detailed in \cite{Maxwell2016}.  To summarize the procedure, the global activation energy, heat release, post-shock flame speed, and the ZND structure \citep{Fickett1979} are first obtained using the Cantera libraries \citep{Goodwin2013} for realistic chemistry with the GRI-3.0 detailed kinetic mechanism \citep{Smith2013}.  The pre-exponential factor, $A$, and diffusion coefficients are then \hl{chosen} such that the one-step model reproduces the correct half reaction length, $\hat{\Delta}_{1/2}$, and also the \hl{correct laminar premixed flame speed at post-shock conditions, prior to auto-ignition,} for a shock travelling at 70\% the theoretical CJ speed for the given quiescent mixture.  \hl{This particular state of reference (70\% CJ) was previously chosen in order to evaluate the laminar flame speed of unburned pockets of gas in} \cref{sec.burningrate}.  The various dimensional and corresponding non-dimensional parameters relevant to one-step combustion model are given in table \ref{tab:params_2dmethane}.

\begin{table*}
  \centering
  \caption{Dimensional and non-dimensional fluid properties, and model parameters, for methane combustion at $\hat{T}_o=300$K and $\hat{p}_o=3500$Pa.}
  {\small
  \resizebox{\columnwidth}{!}{
  \begin{tabular}{llllllllll}
    \hline
    \multicolumn{6}{l}{\textbf{Dimensional properties}} \\
    \hline
    $\hat{\rho_o}$ & 0.04 kg $\textrm{m}^{-3}$    & & $\hat{c_o}$        & 356.36 m $\textrm{s}^{-1}$ & & $\hat{E_a}/\hat{R}$   & 18357.4 K  \\
    $\hat{D}_{CJ}$ & 2243.31 m $\textrm{s}^{-1}$                  & & $\hat{D}_{70\%CJ}$ & 1571.55 m $\textrm{s}^{-1}$ & & $\hat{S}_{L,70\%CJ}$  & 16.39 m $\textrm{s}^{-1}$\\
    $\hat{T}_{CJ}$ & 3159.60 K                    & & $\hat{p}_{CJ}$     & 90.7 kPa    & & $\hat{\rho}_{CJ}$     & 0.07 kg $\textrm{m}^{-3}$  \\
    $\hat{T}_{70\%CJ}$ & 1085.39 K                & & $\hat{p}_{70\%CJ}$ & 81.0 kPa    & & $\hat{\rho}_{70\%CJ}$ & 0.24 kg $\textrm{m}^{-3}$  \\
    $\hat{T}_{VN}$ & 1726.74 K                    & & $\hat{p}_{VN}$     & 169.5 kPa    & & $\hat{\rho}_{VN}$     & 0.32 kg $\textrm{m}^{-3}$ \\
    $\hat{\nu}$    & 1.9x$10^{-4}$ $\textrm{m}^2$ $\textrm{s}^{-1}$ & & $\hat{k}/{(\hat{\rho}\hat{c_p})}$ & 4.8x$10^{-4}$ $\textrm{m}^2 \textrm{s}^{-1}$ & & $\hat{D}$ & 2.0x$10^{-4}$ $\textrm{m}^2$ $\textrm{s}^{-1}$  \\
    $\hat{Q}$             & 6388.0 kJ $\textrm{kg}^{-1}$     & & $\hat{\Delta}_{1/2}$   & 9.68 mm \\

    \hline
    \multicolumn{6}{l}{\textbf{Non-dimensional model parameters}} \\
    \hline
    $\nu$ & 3.68x$10^{-3}$ & & $D_{CJ}$ & 6.30 & & $D_{70\%CJ}$ & 4.41 \\
    $Le$ & 1.32 & & $\Pran$ & 0.709 & & $Sc$ & 0.933 \\
    $\Pran_t$ & 1.0 & & $Sc_t$ & 1.0 & & $\gamma$ & 1.17 \\
    $E_a$ & 45.0 &  & $Q$ & 50.3 & & $A$ & $8.96$x$10^3$  \\
    
  \end{tabular}}
  }
  \label{tab:params_2dmethane}
\end{table*}

\section{Simulation Results}
\label{sec.Sim_results}

\subsection{Preliminary simulation results}

\begin{figure}
  \centering
  \includegraphics[scale=0.24]{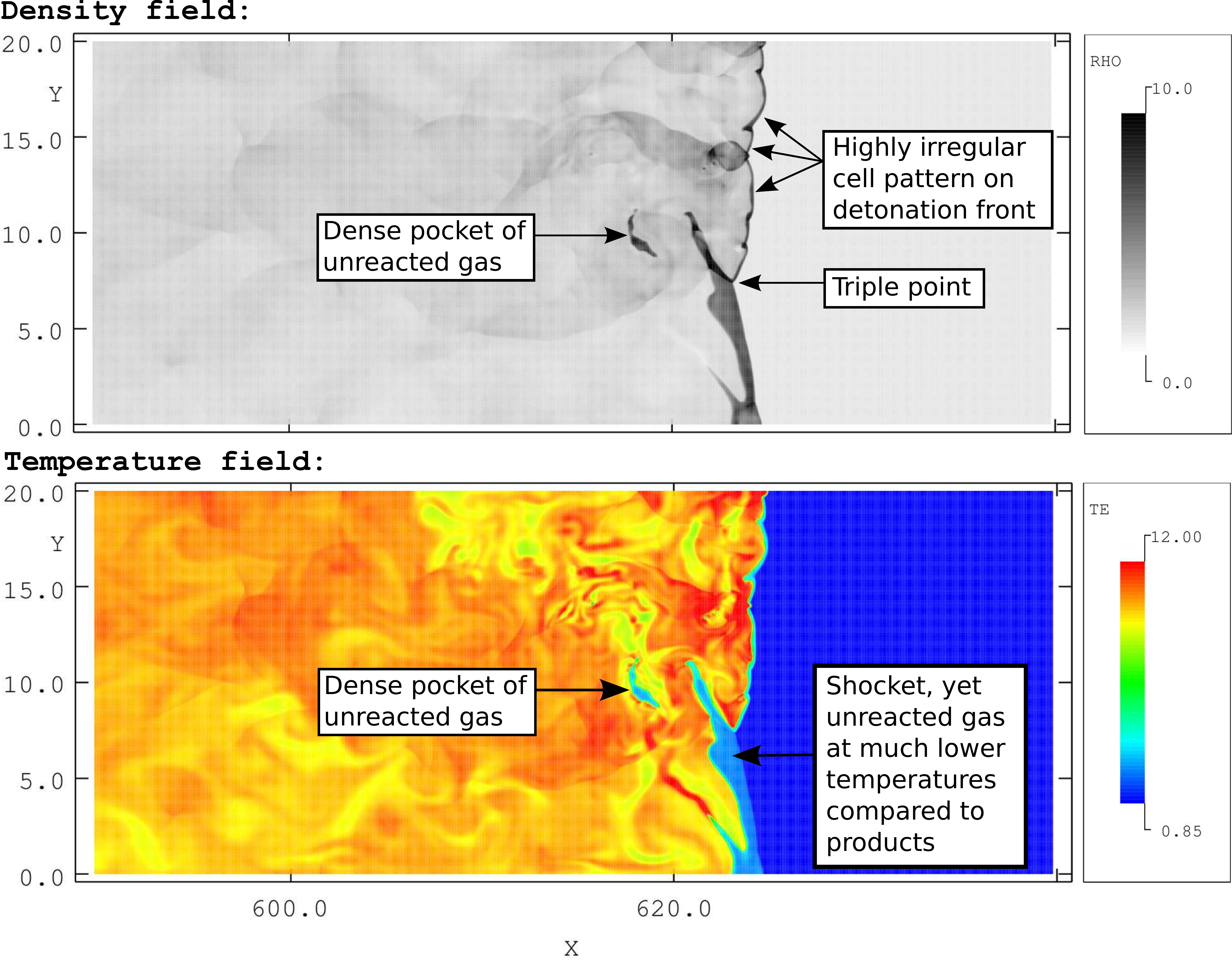}
  \caption[Density and temperature fields for the CLEM-LES with $C_\kappa=1.5$]{Density (top) and temperature (bottom) fields for the CLEM-LES with $C_\kappa=1.5$.  This preliminary result shows the irregular cellular pattern that emerges on the detonation front, and the occasional formation of unburned pockets of reactive gas that appear in the wake.  Note:  density, temperature, and distances are normalized by $\hat{\rho}_o$, $\gamma\hat{T}_o$, and $\hat{\Delta}_{1/2}$, respectively.}
  \label{fig:det_front_ck15}
\end{figure}

\begin{figure}
  \centering
  \includegraphics[scale=0.65]{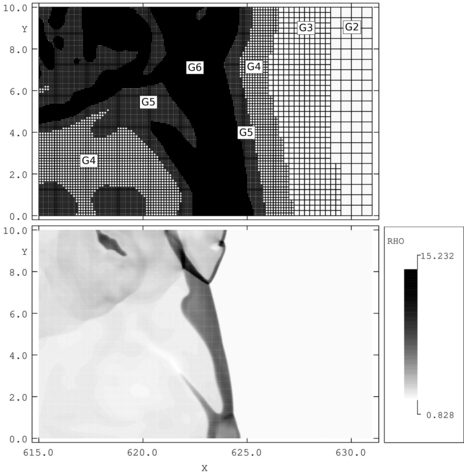}
  \caption[Grid topology for the CLEM-LES with $C_\kappa=1.5$]{Grid topology (top) and corresponding density field (bottom) for a portion of the CLEM-LES simulation shown in figure \ref{fig:det_front_ck15}.  Also shown is the locations of the various grid levels (G2-G6).  Note:  The base grid G1 is always refined to at least grid level G2, everywhere.}
  \label{fig:grid_topology}
\end{figure}

Here, a CLEM-LES simulation was conducted for a $C_\kappa$ value of 1.5 to illustrate the solution.  The maximum resolution of the supergrid was set to $b=1/32$ (32 cells per $\hat{\lambda}_{1/2}$), while the number of subgrid elements within each LES cell was set to $N=16$.  For the AMR implementation, the base grid (G1) has a resolution of $b_{G1}=1$ with 5 additional levels of refinement for a total of 6 grid levels on the supergrid.  Also, the domain height for this simulation was $H=20$, consistent with the experiments in \cref{sec.experiments}.  In the simulation, the wave was allowed to travel approximately $\sim850\hat{\Delta}_{1/2}$ downstream, which was found sufficient to allow the wave to reach a quasi-steady state, in terms of velocity and exhibited cell patterns.  A snapshot of the density and temperature fields of the wave, for a particular instance in time near the end of the channel, are shown in figure \ref{fig:det_front_ck15}.  This particular figure shows that a very irregular pattern emerges on the front.  Clearly, there is one large cell that spans half of the domain height, and several much smaller cells above it.  Also shown in the figure, is the formation of a typical unburned pocket of reactive gas that appears in the wake.  Such pockets are much more cool and dense compared to the surrounding burned products.  Also, to compliment figure \ref{fig:det_front_ck15}, the corresponding grid topology, which indicates the locations of each grid level (G) in a portion of the flow field, is shown in figure \ref{fig:grid_topology} for the same instance in time.  From this figure, it can be verified that the reaction zone is always refined to the finest level, G6 in this case.

\begin{figure}
  \centering
  \includegraphics[scale=0.6]{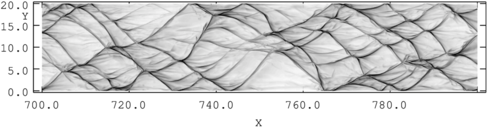}
  \caption[Portion of the numerical soot foil obtained for the CLEM-LES with $C_\kappa=1.5$.]{Portion of the numerical soot foil obtained for the CLEM-LES with $C_\kappa=1.5$.  The streak marks in the figure show the paths of high vorticity associated with triple point trajectories.  Note: Distances are normalized by $\hat{\Delta}_{1/2}$.}
  \label{fig:soot_ck15}
\end{figure}

To gain insight on how these cells evolve as the wave propagates through the channel, a numerical soot foil was obtained by integrating the local vorticity $\Omega(x,y)$ throughout the duration of the simulation, locally at each spatial location, given by
\begin{equation}
\Omega(x,y) = \int_{t=0}^{t}\biggl(\boldsymbol{\nabla}\times\bar{\boldsymbol{u}}(x,y,t)\biggr)\textrm{d}t
\label{eqn.sootFoil}
\end{equation}%
Figure \ref{fig:soot_ck15} shows a portion of the numerical soot foil obtained for a later portion of the channel, from $700<x<800$.  This particular portion is shown as it highlights, in detail, the complex nature of the cellular pattern that emerges.  In this figure, the streak marks follow paths of large vorticity associated with triple point trajectories.  Clearly, in the figure, there are several different cell sizes observed.  These range as large as the domain height, $H$, to cells as small as $\sim H/4$.  Furthermore, there does not appear to be any coherent pattern to the appearance of smaller cells in the presence of larger, more prominent ones.  Also, the triple point paths, given by the streaks on the image, do not follow perfectly straight lines, or at consistent angles.  Sometimes these triple point paths curve more than others in a very irregular manner.  Thus, it is of particular interest to determine how turbulence intensity, a function of $C_\kappa$, might affect the qualitative features discussed above.  Of particular interest is the distance it takes for pockets of unreacted gas to burn up, i.e., the reaction zone thickness.  Also, it is of interest to determine how turbulence affects the overall flow field at the wave front, the formation and burn out of the unreacted pockets of gas, and also the cell patterns that emerge. Capturing such features, numerically, will serve to validate the strategy with experiments in \cref{sec.experiments}, and also those of \cite{Kiyanda2013}.

\begin{figure}
  \centering
  \includegraphics[scale=1.2]{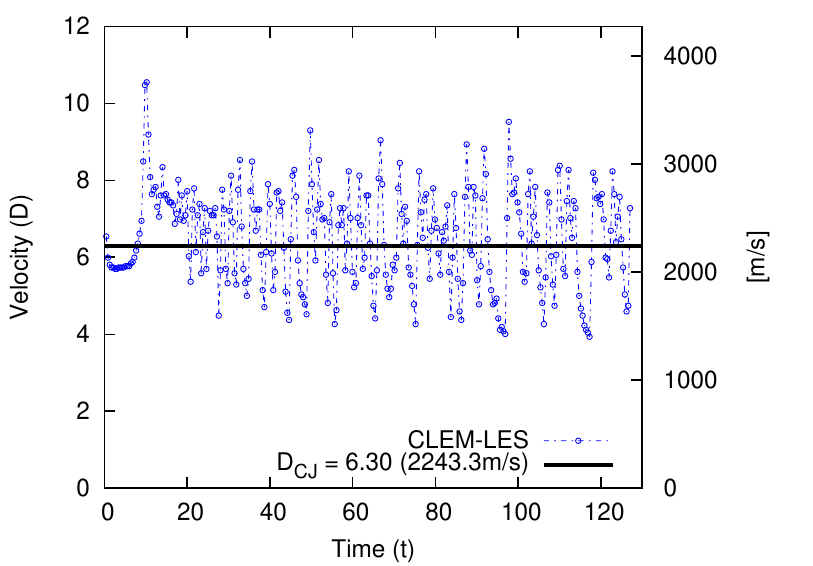}
  \caption[$x$-velocity of detonation for $C_\kappa=1.5$ and $b=1/32$ as a function of time.]{$x$-velocity of detonation for $C_\kappa=1.5$ and $b=1/32$ as a function of time, measured along bottom wall at $y=0$.  Note: $D$ and $t$ are normalized by $\hat{D}_{CJ}$ and ($\hat{\Delta}_{1/2}/\hat{c}_{o}$), respectively.}
  \label{fig:Dspeed_15}
\end{figure}

Finally, figure \ref{fig:Dspeed_15} shows the $x$-velocity the detonation wave as a function of time, measured along bottom wall at $y=0$.  Initially, the wave is clearly over-driven above the theoretical CJ-value.  However, after $t>55$, sufficient time has passed to allow the detonation wave to decelerate from the over-driven state and reach velocities closer to the CJ-speed.  In fact, beyond $t>55$, the averaged speed of the wave was $D=6.35$ \hl{(2260$\textrm{ms}^{-1}$)}, within $1\%$ error of the CJ-value of $D_{CJ}=6.30$ \hl{(2240$\textrm{ms}^{-1}$)}.  Clearly, in figure \ref{fig:Dspeed_15}, the velocity peaks and valleys also appear very irregular in nature, with no coherent pattern.  These velocity measurements correspond to cells which form along the wall.  In fact, the maxima correspond to the start of each cell cycle observed along the bottom wall, where triple point collisions occur.  The minima represent the end of the cell cycle where the velocity has decayed below the CJ-value.  Although the cell cycles appear random, much like the cells shown in the numerical soot foil of figure \ref{fig:soot_ck15}, it is of particular interest to determine how the wave velocities are statistically distributed on the wave front, given the turbulence intensity ($C_\kappa$).  Furthermore, it is of prime importance to compare such a velocity distribution with available experimental data obtained in this study and also in \cite{Kiyanda2013}, for the purpose of CLEM-LES validation.

\subsection{Grid convergence study}
\label{sec.Detonation_gridConvergence}

Prior to investigating the effect of turbulent mixing rates on the detonation propagation characteristics, it is important to first determine a sufficient resolution which resolves both the reaction rate, and qualitative features of the cellular structure.  \hl{For this resolution study, the domain height is kept constant at $H=20$.  Also, $C_\kappa=1.5$ is specified for the CLEM-LES simulations, and the number of subgrid elements within each LES cell is held constant at $N=16$.  Only the super-grid resolution, $b$, is varied.  This value for $N$ was found, in previous work} \citep{Maxwell2015}, \hl{to be sufficient at capturing stirring events and eddy-flame interactions on the subgrid, while optimizing computational efficiency associated with the method.  Also, it is worth noting here that at $b=1/32$ resolution with $N=16$, the laminar flame speed is fully resolved in one-dimension, see} \cite{Maxwell2015} \hl{for details.} 

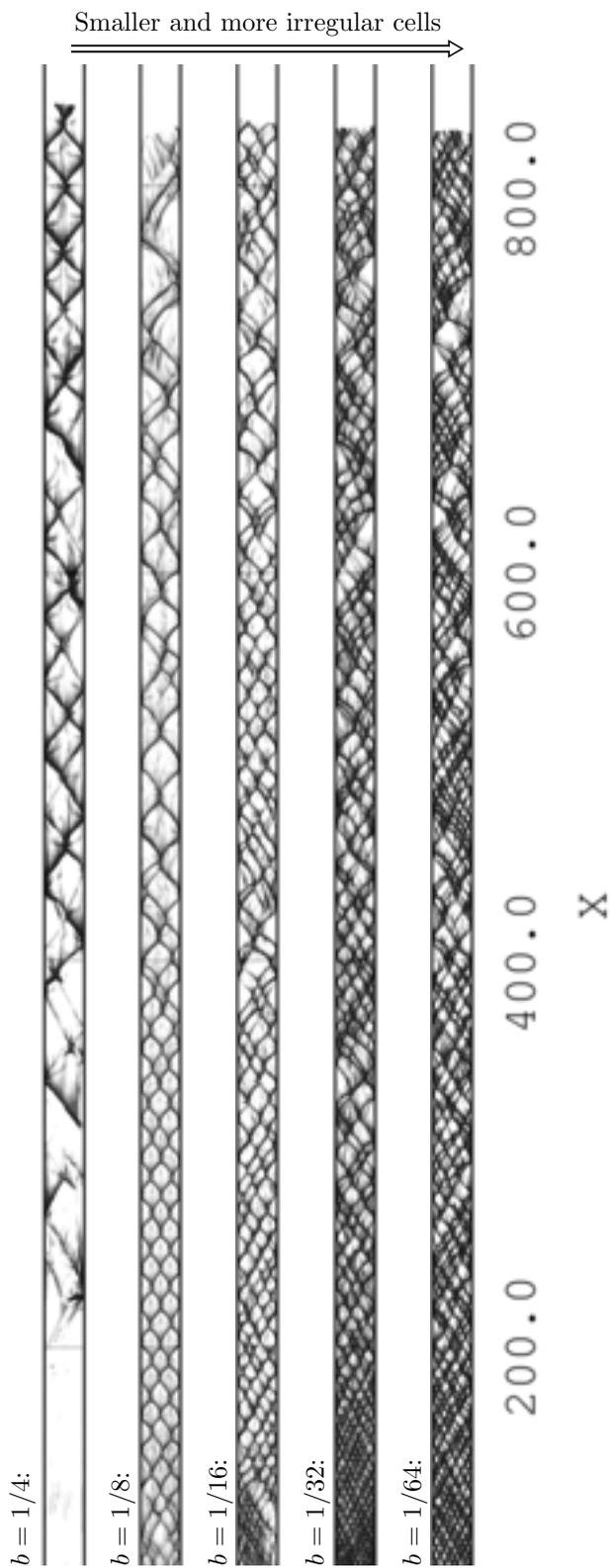
\begin{sidewaysfigure}
  \vspace{15cm}
  \raggedright $b=1/4$: \\
  \includegraphics[trim=5.5cm 7cm 6.5cm 9cm, clip=true, scale=1.0]{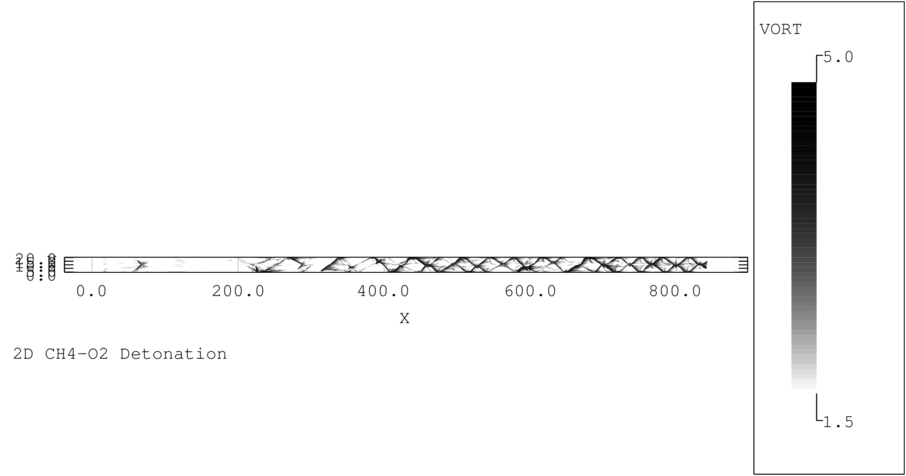} \\
  $b=1/8$: \\
  \includegraphics[trim=5.5cm 7cm 6.5cm 9cm, clip=true, scale=1.0]{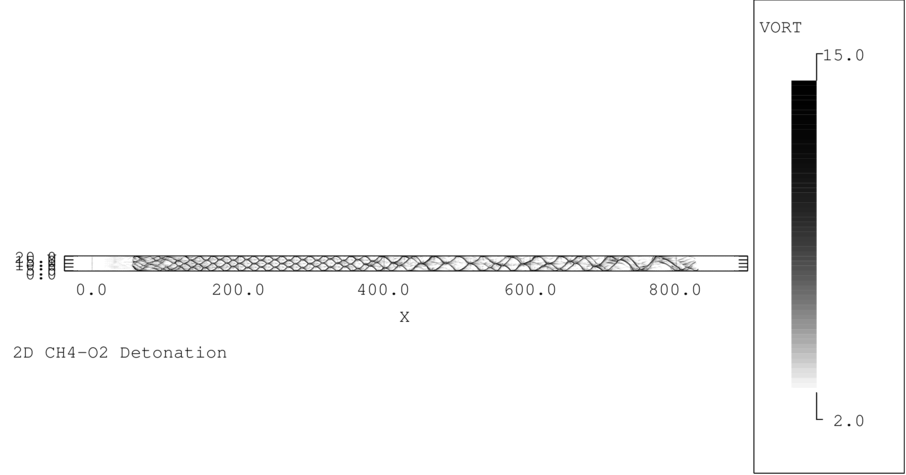} \\
  $b=1/16$: \\
  \includegraphics[trim=5.5cm 7cm 6.5cm 9cm, clip=true, scale=1.0]{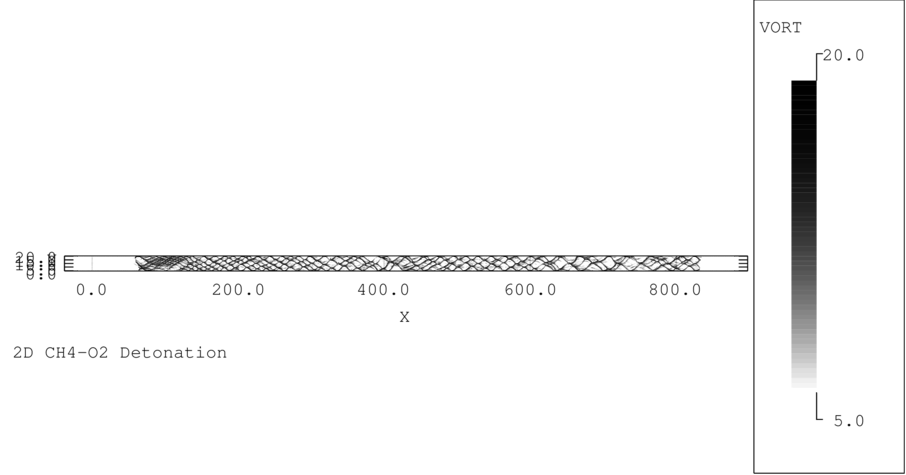} \\
  $b=1/32$: \\
  \includegraphics[trim=5.5cm 7cm 6.5cm 9cm, clip=true, scale=1.0]{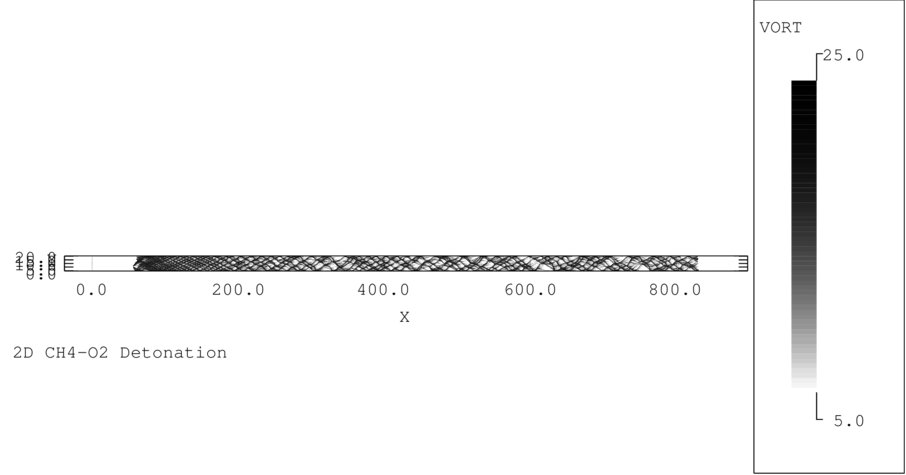} \\
  $b=1/64$: \\
  \includegraphics[trim=5.5cm 5.0cm 6.5cm 9cm, clip=true, scale=1.0]{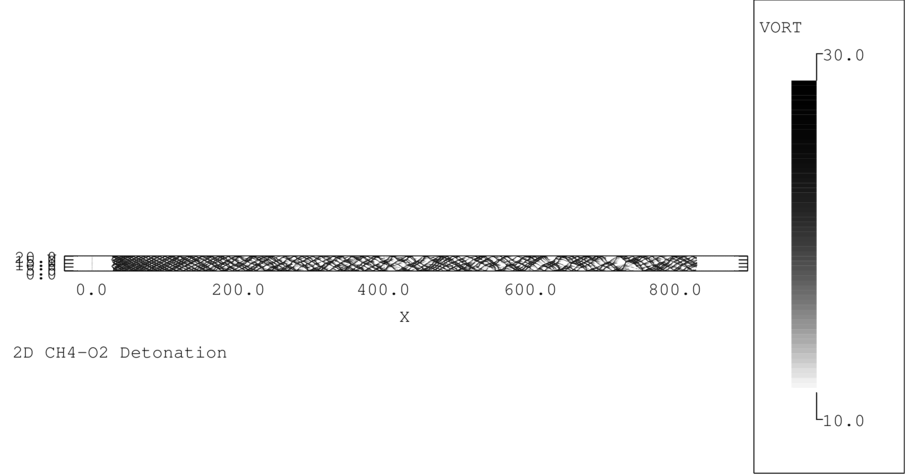}
  \caption[Numerical soot foils for $C_\kappa=1.5$ at various resolutions.]{Numerical soot foils for $C_\kappa=1.5$ at various resolutions, $b$ (all with $N=16$ elements per LES cell).  \\ Note:  Distance $x$ is normalized by $\hat{\Delta}_{1/2}$.}
  \begin{tikzpicture}[remember picture,overlay]
    \node[anchor=south] at ($(current page.south) + (22.3,-1.75)$) (A) {};
    \node[anchor=south] at ($(current page.south) + (22.3,-7.25)$) (B) {};
    \draw[vecArrow] (A) to (B);
    \node[anchor=south,rotate=-90] at ($(current page.south) + (22.35,-4.25)$) (C) {Smaller and more irregular cells};
  \end{tikzpicture}
  \label{fig:res_openshutters}
\end{sidewaysfigure}

First, to assess the qualitative behaviour of resolution on the inherent cellular instability of the wave front, numerical soot foils were obtained for each simulation and compared in figure \ref{fig:res_openshutters}.  For all resolutions, the simulations are run up to $t=127.5$.  As observed in figure \ref{fig:res_openshutters}, the lowest resolution, $b=1/4$, produces only large cells whose size appears to be mode locked by the channel height.  Furthermore, the cells at this resolution exhibit a very regular self-repeating pattern. As the resolution is increased to $b=1/8$, the detonation wave continues to exhibit large cells on the order of the domain height.  However, the cells at this resolution appear to be slightly irregular in size and shape.  Also, the appearance of smaller cells within the larger cell structure are observed starting around $x\sim550$.  For $b=1/16$, the cell sizes appear much smaller and even more irregular.  Also, there appears to be much more variation in cell sizes observed throughout the channel.  For example, at $x\sim200$ there are approximately 4 cells spanning the domain height.  At $x\sim400$, the cells are intermittently much larger, with about 1 cell spanning the domain height.  At $x\sim500$ there are 2 cells across the domain height.  At $x\sim700$, the cell size increases to 1 cell across the domain height.  Then, at $x\sim800$ there are once again approximately 2 cells spanning the domain height.  For the higher resolutions, $b=1/32$ and $b=1/64$, the same stochastic cell size behaviour is observed as the $b=1/8$ case.  For these two higher resolution cases, however, the pattern appears even more irregular than the $b=1/8$ case.  In fact, the $b=1/32$ and $b=1/64$ cases are qualitatively similar to each other in terms of observed cell sizes, pattern behaviour, and irregularity.

To effectively quantify the average rate at which reactant was consumed behind the leading shock wave, at any given instant in time, the average distance for the reactant to be consumed behind the leading shock wave was measured.  This was indicative of the time it takes for the reactant burn up behind the leading shock.  To measure this average \emph{reaction zone thickness}, a Favre-averaging technique in spatial slices on the super-grid, $dx$, was applied to give a mass weighted average reactant and density profile along the channel length for any given time.  This procedure was previously applied in \cite{Radulescu2007b}.  Formally, the Favre-average reactant profile is given by the expression
\begin{equation}
\tilde{Y}(x,t) = \frac{\int_{y=0}^{H}\bar{\rho}(x,y,t)\tilde{Y}(x,y,t)\textrm{d}y}{\int_{y=0}^{H}\bar{\rho}(x,y,t)\textrm{d}y}
\label{eqn.Favre_space}
\end{equation}%
where the average density profile, in $x$, is found by
\begin{equation}
\bar{\rho}(x,t) = \frac{\int_{y=0}^{H}\bar{\rho}(x,y,t)\textrm{d}y}{H}
\label{eqn.AvgDensity_space}
\end{equation}%
Finally, the Favre-averaged mass fractions are ensemble averaged for $k=50$ random instances in time.  Thus,
\begin{equation}
\tilde{Y}(x-x_s) = \frac{\sum_{i=0}^{k}\tilde{Y}(x-x_s,t_i)}{k}
\label{eqn.ensemble_time}
\end{equation}%
Here $x_s$ was the location of the shock wave determined by the location where the right-most maximum gradients in the spatially averaged density occur.  Also, the sample slices for which averaging is done is taken as the minimum super-grid resolution size, $dx=b$. Finally, it should be noted that this ensemble averaging only considers profiles when $t>40$.  This was done to allow sufficient time for the detonation wave to decelerate from the over-driven state and reach its self-sustained equilibrium structure.

\begin{figure}
  \centering
  a)\includegraphics[scale=1.2]{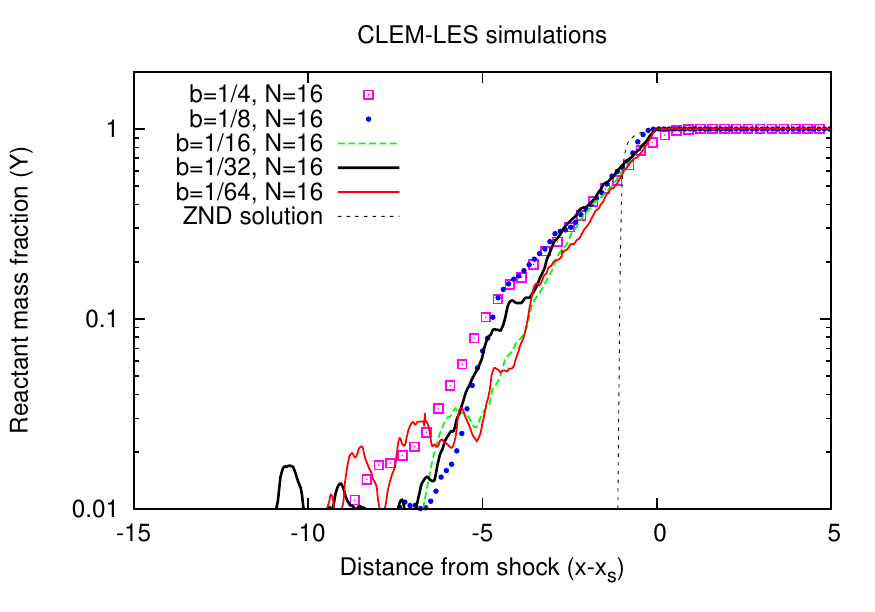}\\
  b)\includegraphics[scale=1.2]{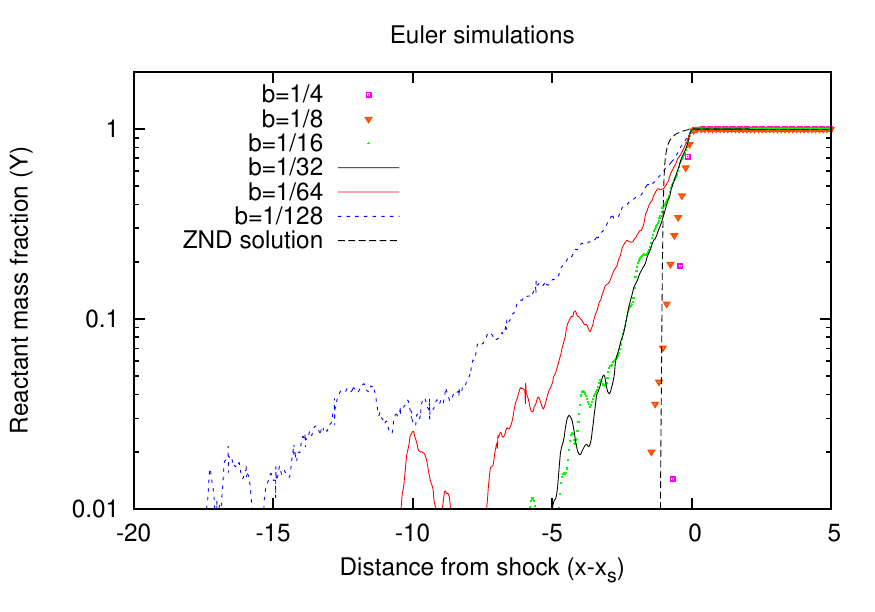}  
  \caption[Effect of resolution on average reactant profiles for a) the CLEM-LES and compared to b) Euler methods.]{Effect of resolution on average reactant profiles for a) the CLEM-LES and compared to b) Euler methods. Note: Distances $x$ and $x_s$ are normalized by $\hat{\Delta}_{1/2}$.}
  \label{fig:res_profiles}
\end{figure}

\begin{figure}
  \centering
  \includegraphics[scale=1.2]{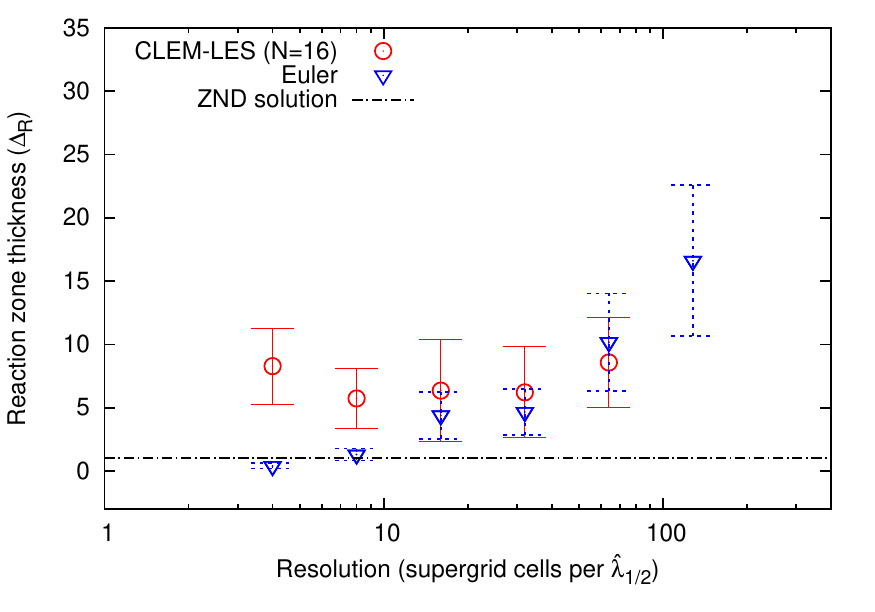}
  \caption[Effect of resolution on reaction zone thickness ($\Delta_R$) for CLEM-LES and compared to Euler methods.]{Effect of resolution on the mean reaction zone thickness ($\Delta_R$) for CLEM-LES and compared to Euler methods. The error bars indicate the standard deviation in $\Delta_R$ for each simulation.  Note: $\Delta_R$ is normalized by $\hat{\Delta}_{1/2}$.}
  \label{fig:res_thickness}
\end{figure}

In figure \ref{fig:res_profiles}a, the average profiles obtained for $\tilde{Y}(x-x_s)$ are presented for the CLEM-LES at various resolutions ($b$). For the CLEM-LES, it can clearly be seen that the average rate of depletion of $\tilde{Y}(x-x_s)$ behind the shock wave {does not vary significantly} across the range of resolutions tested.  In fact, all of the \hl{simulations} appear to consume 90\% of the reactant within the same average distance, $\sim5\hat{\Delta}_{1/2}$.  Finally, only the highest resolutions, $b=1/32$ and $b=1/64$, develop fluctuations in reactant mass fractions when $\tilde{Y}(x-x_s)<10\%$, and thus has a slightly lengthened structure compared to lower resolution cases.  These fluctuations are believed to arise due to unresolved sampling of high frequency statistics sufficiently far from the leading shock wave. For the low resolution cases, even though the same reaction zone thickness is captured, these high frequency instabilities are effectively filtered out.  This suggests that the reaction rate, in general, is not very sensitive to resolution.  However, the irregular propagation behaviour and fine scale qualitative observations are sensitive to resolution. To compare the performance of the CLEM-LES with the traditional use of Euler simulations, figure \ref{fig:res_profiles}b shows the average reactant profiles for $\tilde{Y}(x-x_s)$ using the Euler method across the range of resolutions from $b=1/4$ to $b=1/128$.  \hl{It should be noted here that the Euler simulations adopt the same second order accurate exact Godunov solver which was applied to the advection terms on the CLEM-LES supergrid} \citep{Falle1991}.  Clearly, the Euler method does not provide any convergence of solution with increased resolution.  This can be observed by the lengthening of the structure, or distance it takes for reactant to become consumed, as the resolution increases.  To quantify the reaction zone thickness of the structure of each simulation, the thickness, $\Delta_R$, is arbitrarily taken as the distance from shock location, at $x=x_s$, to the position where $\tilde{Y}(x-x_s)<2\%$.  Figure \ref{fig:res_thickness} shows that across the range of resolutions, the average reaction zone thickness for the CLEM-LES is not very sensitive to resolution, whereas the Euler method clearly shows an increase in average thickness as resolution increases.  Figure \ref{fig:res_thickness} also shows the range of reaction zone thickness observed throughout the random sampling process.  The error bars thus serve as the variation in thickness for the 50 random samples used in the ensemble averaging process of each simulation.  Clearly, the Euler method yields an increasing variation in thickness as the resolution is increased.  Also at high resolutions, $b\ge1/64$, the Euler method yields larger, and more stochastic, range in thickness compared to the CLEM-LES.  The CLEM-LES, on the other hand, appears to converge to a statistically consistent range of thickness observed with each random sample.  This can be seen for $b=1/32$ and $b=1/64$ where the reaction zone thickness is found to vary stochastically between $\Delta_R=3-16\hat{\Delta}_{1/2}$ for both resolutions.  Furthermore, the average thickness for the CLEM-LES remains consistent around $\Delta_R=6-9\hat{\Delta}_{1/2}$ for all resolutions.

To summarize the findings in this section, a resolution of $b=1/32$ has been demonstrated to sufficiently resolve and capture the structure size,  high frequency instabilities exhibited through cell patterns, and overall propagation behaviour.  Thus, $b=1/32$ is the resolution used throughout the remainder of the paper.

\clearpage

\subsection{Effect of the Kolmogorov parameter ($C_k$)}
\label{sec.CK_effect}

In order to investigate the role of turbulent mixing rates on the detonation front cellular patterns, burning rates, and irregularity, the constant $C_\kappa$ has been varied.  This was done using a maximum supergrid resolution of $b=1/32$ with $N=16$ subgrid elements within each supergrid cell.  As before, to achieve this resolution on the supergrid, the base grid (G1) has a resolution of $b_{G1}=1$ with 5 additional levels of refinement.  It should also be noted that a formal grid resolution study for each $C_\kappa$ was not conducted here.  In the previous section, this particular resolution and AMR configuration was found adequate to resolve the qualitative features associated with cell sizes and patterns, and irregularity.  Furthermore, this resolution was found to give converged burning rates behind the shock, which was measured quantitatively by considering the size of the observed reaction zone thickness ($\Delta_R$).  Also, the one-dimensional laminar flame speed is resolved at this resolution.  As discussed previously, in \cref{sec.les_eqns_super}, the turbulent viscosity and dissipation rates, $\nu_t$ and $\epsilon$, are both functions of $C_\kappa$, which are consequently functions of $k^{sgs}$.  Thus turbulent mixing rates, and hence combustion rates, are expected to be influenced by changes in $C_\kappa$.  For the numerical experiment in this section, $C_\kappa$ have been varied from 1.2 to 10.0.  As was done in the previous section, the qualitative features are compared using numerically obtained soot foil images.  The domain height for all simulations in this section have once again been kept constant at $H=20$.  Also, the average reaction zone thickness has been collected for each simulation.

\begin{sidewaysfigure}
 \vspace{15cm}

  \raggedright Euler: \\
  \includegraphics[trim=5.5cm 7cm 6.5cm 9cm, clip=true, scale=1.0]{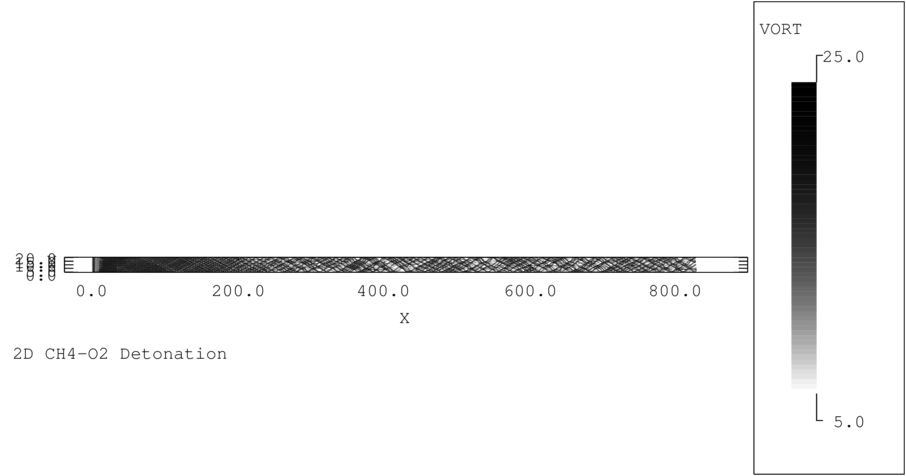} \\
  $C_\kappa=1.2$: \\
  \includegraphics[trim=5.5cm 7cm 6.5cm 9cm, clip=true, scale=1.0]{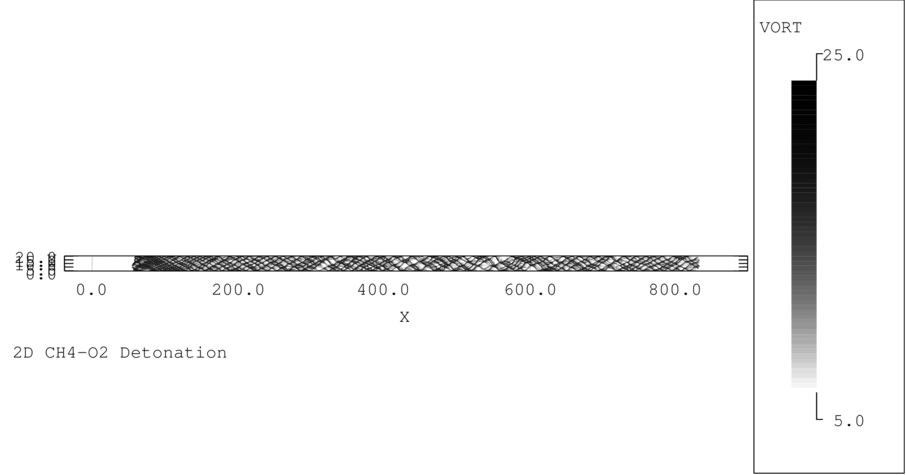} \\
  $C_\kappa=1.5$: \\
  \includegraphics[trim=5.5cm 7cm 6.5cm 9cm, clip=true, scale=1.0]{Figure12d.png} \\
  $C_\kappa=3.0$: \\
  \includegraphics[trim=5.5cm 7cm 6.5cm 9cm, clip=true, scale=1.0]{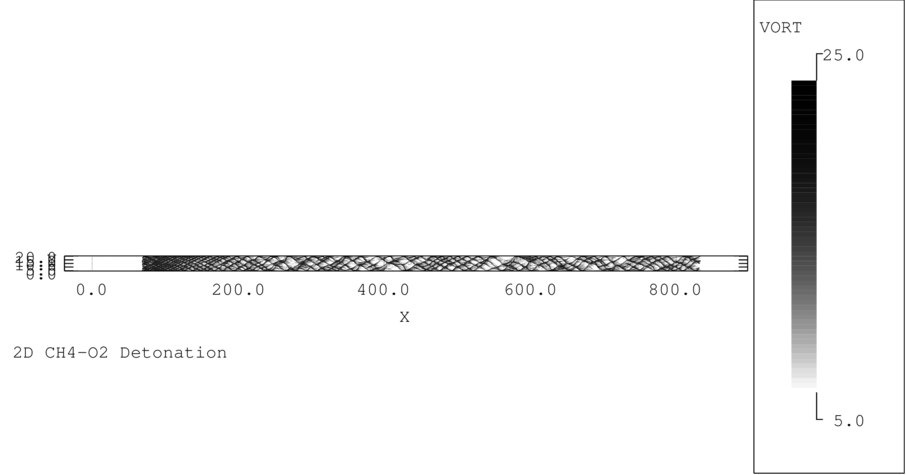} \\
  $C_\kappa=4.0$: \\
    \includegraphics[trim=5.5cm 7cm 6.5cm 9cm, clip=true, scale=1.0]{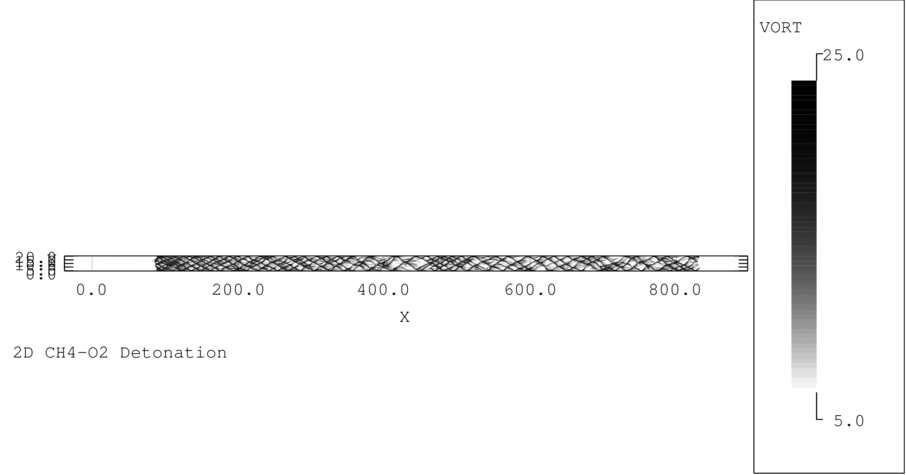} \\
  $C_\kappa=6.7$: \\
    \includegraphics[trim=5.5cm 7cm 6.5cm 9cm, clip=true, scale=1.0]{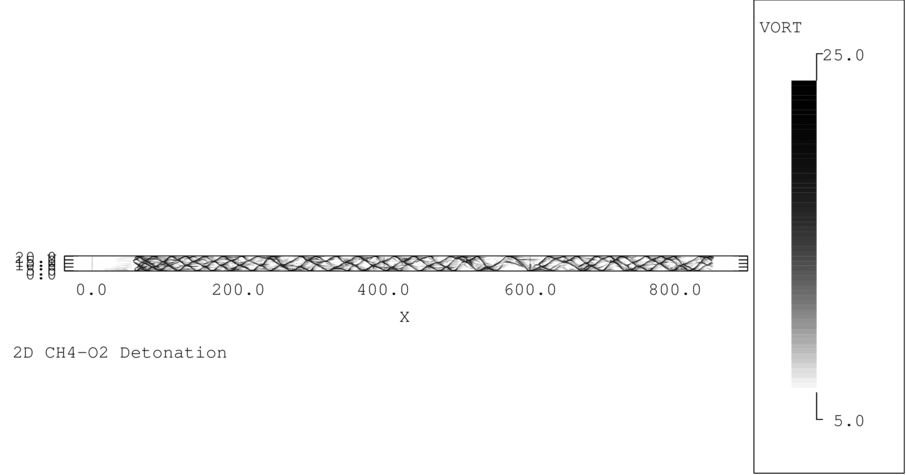} \\
  $C_\kappa=10.0$: \\
  \includegraphics[trim=5.5cm 5cm 6.5cm 9cm, clip=true, scale=1.0]{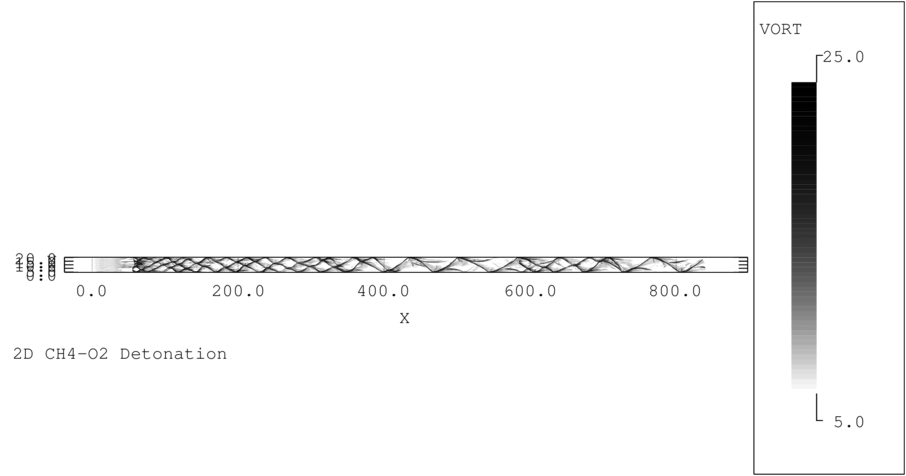}

  \caption[Numerical soot foils for various $C_\kappa$ values and an Euler simulation.]{Numerical soot foils for various $C_\kappa$ values and an Euler simulation (all with resolution of $b=1/32$ and, for the LES, a further $N=16$ elements per supergrid cell). Note:  Distance $x$ is normalized by $\hat{\Delta}_{1/2}$.}
   \begin{tikzpicture}[remember picture,overlay]
     \node[anchor=south] at ($(current page.south) + (22.3,-1)$) (A) {};
     \node[anchor=south] at ($(current page.south) + (22.3,-8)$) (B) {};
     \draw[vecArrow] (A) to (B);
     \node[anchor=south,rotate=-90] at ($(current page.south) + (22.35,-5.0)$) (C) {Larger cells};
   \end{tikzpicture}

  \label{fig:ck_openshutters}
\end{sidewaysfigure}

First, numerically obtained soot foils are shown in figure \ref{fig:ck_openshutters} for the range of $C_\kappa$ values investigated (1.2-10).  Also, the CLEM-LES soot foil images are compared against the Euler simulation, which also has a resolution of 32 cells per half-reaction length.  \hl{Clearly the Euler simulation exhibits much smaller cells compared to all of the CLEM-LES soot foil images.}  For the CLEM-LES at $C_\kappa=1.2$, in figure \ref{fig:ck_openshutters}, the cells are still fairly small compared to the channel height, however there is some irregularity to the pattern observed.  As $C_\kappa$ is increased, the observed cell sizes also increase.  For $C_\kappa=3.0-4.0$ it is observed that the range of cell sizes span from 3 cells to 1 cell per channel height, intermittently, throughout the simulations.  As $C_\kappa$ is increased beyond 6.7, the cells become even larger, with an observed range of 2 cells to 1/2 cell per channel height.

\begin{figure}
  \centering
  \includegraphics[scale=1.2]{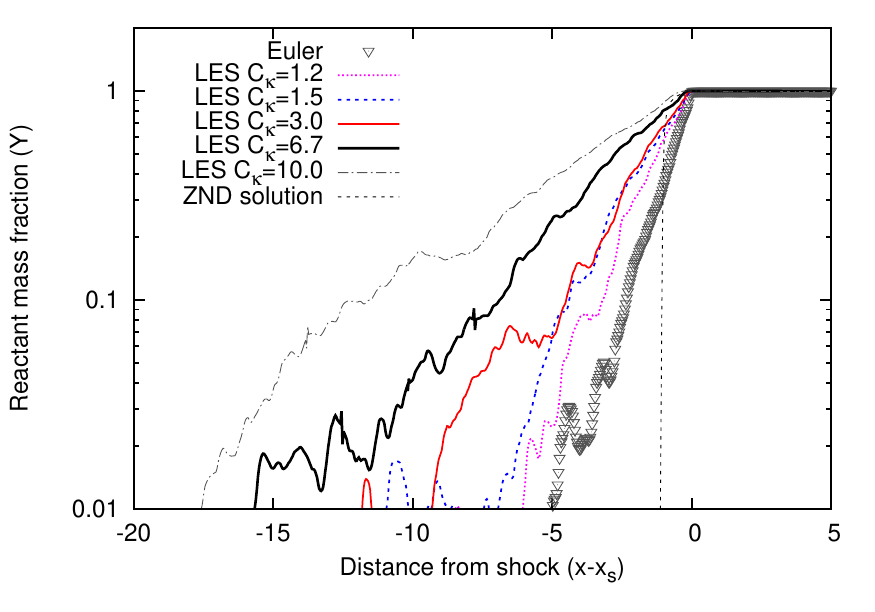}
  \caption[Effect of turbulent intensity (through $C_\kappa$) on average reactant profiles for the CLEM-LES.]{Effect of turbulent intensity (through $C_\kappa$) on average reactant profiles for the CLEM-LES (all with resolution $b=1/32$ and $N=16$ elements per LES cell).  Also shown is the average reactant profile obtained from the Euler method at $b=1/32$ resolution. Note: Distances $x$ and $x_s$ are normalized by $\hat{\Delta}_{1/2}$.}
  \label{fig:ck_profiles}
\end{figure}

Next, in figure \ref{fig:ck_profiles}, the average reactant profiles are compared for the range of $C_\kappa$ values using the procedure described in the previous section.  Also, the simulations are compared to the Euler simulation with the same resolution as the CLEM-LES supergrid.  Clearly, as $C_\kappa$ increases, not only does the cell size increase, but the average reaction zone thickness of the detonation wave also increases.  This observation on reaction zone thickness is made by considering the distance it takes for $\tilde{Y}(x-x_s)\rightarrow0$.  Furthermore, this relation between the cell size, and the reaction zone thickness is consistent with \hl{earlier correlations, as reported by} \cite{Lee1984}.  In this sense, as $C_\kappa$ increases, $\Delta_R$ increases, which consequently exhibits larger cell sizes.  On the other hand, the Euler method at the same resolution, which is subject to a large amount of numerical diffusion, burns fuel at a much quicker rate than all of the CLEM-LES simulations. \hl{It is noted that although the Euler simulations and CLEM-LES have the same numerical diffusion on the supergrid, where the pressure evolution is solved for both strategies, the CLEM-LES has much better closure of the reaction rate since the higher resolution subgrid itself has much less numerical diffusion that contribute to mixing and chemical reaction.} Finally, due to the sensitivity of the reaction zone thickness ($\Delta_R$), and consequently the cell size, on turbulent fluctuations, this permits us to fit the Kolmogorov parameter ($C_\kappa$) against experimental observations.

\section{Discussion}

\label{sec.propagation_discussion}

\subsection{Qualitative comparison of the flow evolution with experiments}
\label{sec.qualitycompare}

\begin{figure}
  \centering
  \includegraphics[scale=0.5]{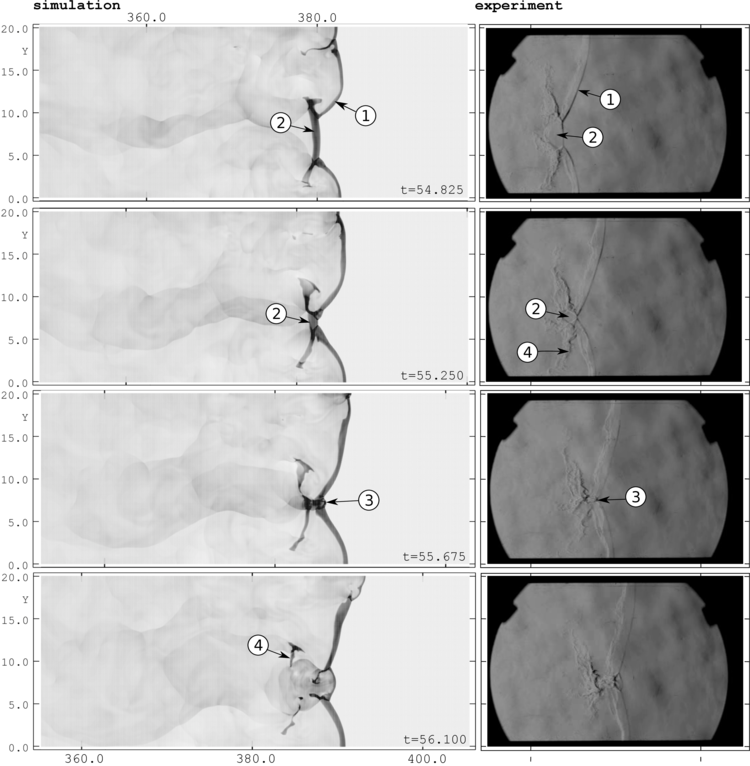}
  \caption[Numerical density evolution for $C_\kappa=6.7$ and the corresponding experimental Schlieren photography (this study) of a detonation triple point collision process (part 1 of 2).]{Numerical density evolution for $C_\kappa=6.7$ (left) and the corresponding experimental Schlieren photography (this study - right) of a detonation triple point collision process.  The time in between images presented above is $\Delta t=0.425$ (11.53\gmu s).  The remaining sequence of images are continued in figure \ref{fig.evolution_cont}.}
  \label{fig.evolution}
\end{figure}

\begin{figure}
  \centering
  \includegraphics[scale=0.5]{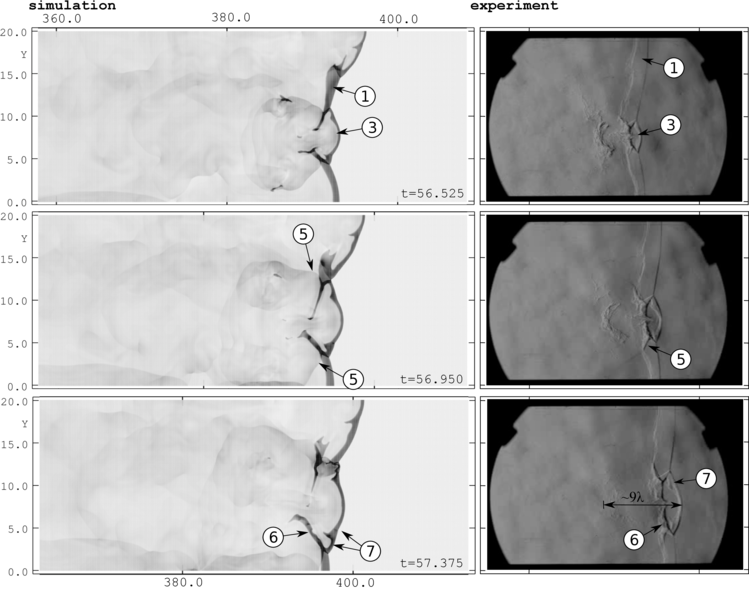}
  \caption[Numerical density evolution for $C_\kappa=6.7$ and the corresponding experimental Schlieren photography \citep{Bhattacharjee2013b} of a detonation triple point collision process (part 2 of 2).]{Continuation of figure \ref{fig.evolution}. Numerical density evolution for $C_\kappa=6.7$ (left) and the corresponding experimental Schlieren photography (right) of a detonation triple point collision process.}
  \label{fig.evolution_cont}
\end{figure}

\begin{figure}
  \centering
  \includegraphics[scale=0.6]{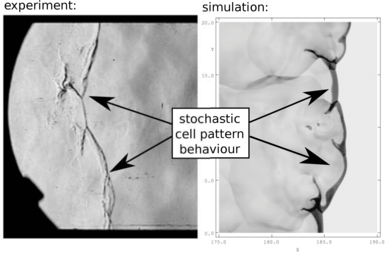}
  \caption[Comparison of experiments and the CLEM-LES with $C_\kappa=6.7$, both showing stochastic behaviour.]{Comparison of experiments and the CLEM-LES with $C_\kappa=6.7$, both showing an instant where stochastic behaviour is observed such that a different cell pattern is observed compared to figure \ref{fig.evolution}.}
  \label{fig:stochastic_behaviour}
\end{figure}

A visual comparison of a flow field, obtained through resolved LES and the corresponding experiment of \cref{sec.experiments}, is shown in figures \ref{fig.evolution} and \ref{fig.evolution_cont} for an instance where a triple-point collision occurs for a detonation whose cell size is comparable to the channel height.  For the LES, the density evolution is shown for various instances of the collision process for $C_\kappa=6.7$.  This value of $C_\kappa$ was found to produce cell sizes and overall qualitative behaviour comparable to experimental observation. Here, the channel height is kept at $H=20$.  For the experiment, Schlieren images of each corresponding instance are obtained from high speed photography and are also shown in the figures. For comparison, many features are noted in the figures (features 1-7).  In figure \ref{fig.evolution}, the first feature (1), is a Mach shock with a turbulent reaction zone that follows closely.  In both the simulation, and the experiment, the reaction zone is slightly decoupled from the shock owing to local unsteadiness of the wave front.  This can be seen by the thickening of the shocked unburned gas region as time evolves.  In the LES, however, this wave appears to be more irregular, containing what appears to be a smaller and less pronounced cell structure (cells within cells).  It should be noted that experimental observations, of \cref{sec.experiments}, report a very stochastic nature of the propagating wave.  In this sense, random bifurcations on wave fronts and changes in cell patterns are expected.  In some cases much smaller cells are observed experimentally and numerically for the same quiescent conditions and numerical parameters, see figure \ref{fig:stochastic_behaviour}.  Feature (2), in figure \ref{fig.evolution}, is a completely decoupled incident shock-reaction zone, as can be seen by the large region of dense unburned gas.  As the two triple-points approach each other, transverse shock waves compress this gas further.  At the focal point of the collision, owing to the increased pressure and temperature, the unburned pocket (2) ignites very rapidly.  This ``explosion", noted by feature (3), drives transverse shock waves and a strong Mach shock forward coupled with very rapid burning.  Feature (4) is a pocket of unburned gas which is not ignited by compression associated with the original Mach shocks (1).  Instead, these pockets mix and burn, \hl{on their surfaces}, along turbulent shear layers. Due to shock compression associated with the passage of transverse shock waves, these shear layers experience increased burning rates through turbulent mixing of the burned and unburned gases via Richtmyer-Meshkov instability.  These transverse shocks, labelled as feature (5) in figure \ref{fig.evolution_cont}, originate from feature (3).  Feature (6), in figure \ref{fig.evolution_cont}, is a new pocket of unburned gas that forms along slip lines associated with the collision process.  In fact, the formation of such pockets, in this manner, is typical of irregular detonation propagation \citep{Subbotin1975,Kiyanda2013,Austin2003,Radulescu2005,Oran1982,Radulescu2007b}.  Finally, feature (7), in figure \ref{fig.evolution_cont}, is a bifurcation observed on the Mach shock of feature (3), owing to the unstable nature of methane-air detonations.  This bifurcation also occurs in the experiment but is less pronounced.  Although features (1-7) are captured by the LES, a key difference is observed in the burning rate behind behind the Mach shock originating from feature (3).  In the simulation, the size of this region grows slightly faster compared to experiment, \hl{owing to the lower detonation velocity reported in} \cref{sec.expvelocity}.

\begin{figure}
  \centering
  \includegraphics[scale=0.8]{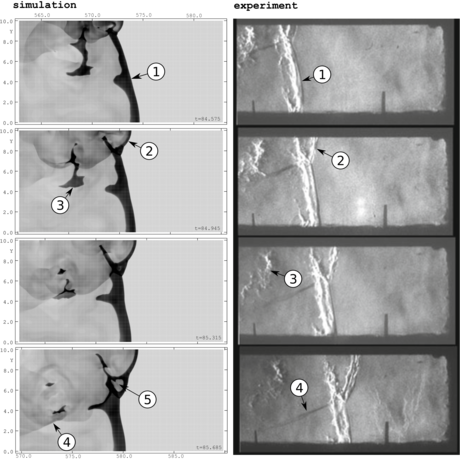}
  \caption[Numerical density evolution for $C_\kappa=6.7$ and the corresponding experimental Schlieren photography \citep{Kiyanda2013} of an irregular detonation propagation in a channel (part 1 of 2).]{Numerical density evolution for $C_\kappa=6.7$ (left) and the corresponding experimental Schlieren photography \citep{Kiyanda2013} (right) of an irregular detonation propagation in a channel. The time in between images presented above is $\Delta t=0.37$ (10.0\gmu s). The remaining sequence of images are continued in figure \ref{fig.evolution2_cont}.}
  \label{fig.evolution2}
\end{figure}

\begin{figure}
  \centering
  \includegraphics[scale=0.8]{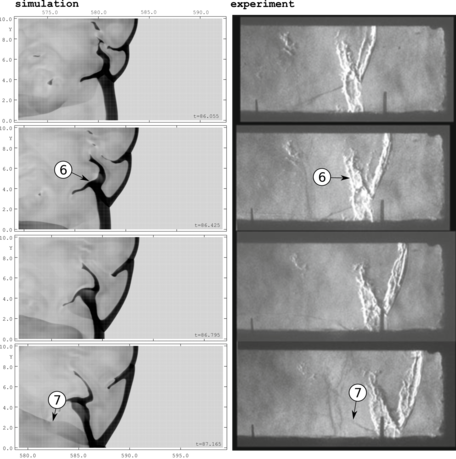}
  \caption[Numerical density evolution for $C_\kappa=6.7$ and the corresponding experimental Schlieren photography \citep{Kiyanda2013} of an irregular detonation propagation in a channel (part 2 of 2).]{Continuation of figure \ref{fig.evolution2}. Numerical density evolution for $C_\kappa=6.7$ (left) and the corresponding experimental Schlieren photography \citep{Kiyanda2013} (right) of an irregular detonation propagation in a channel. The time in between images presented above is $\Delta t=0.37$ (10.0\gmu s).}
  \label{fig.evolution2_cont}
\end{figure}

To further compare the performance of the LES with available experimental data \citep{Kiyanda2013}, the simulation was repeated with a channel height of $H=10$.  Both the density evolution from the numerical simulation, and the corresponding instances obtained in the experiment through Schlieren photography are shown in figures \ref{fig.evolution2} and \ref{fig.evolution2_cont}.  In this case, all model parameters, including $C_\kappa$, are consistent with the previous experiment, except the reduced channel height permits only half a cell to form.  Thus the cell pattern is effectively influenced by the channel walls, or \emph{mode-locked} into a cell size which corresponds roughly to \hl{twice} the channel height.  In this case, only a single transverse wave is observed.  Again, for qualitative comparison, many features are noted in the figures (features 1-7).  Feature (1), in figure \ref{fig.evolution2}, is the incident shock wave and de-coupled reaction zone, as observed by the region of very dense, shocked, and unburned fuel following the shock.  From the top wall of the channel, Feature (2) is a Mach stem, which initially results from a transverse wave collision with the wall.  As observed in figure \ref{fig.evolution}, \hl{this Mach stem travels faster than the incident wave, owing to the wave collision process.}  Feature (3), in figure \ref{fig.evolution2}, is a pocket of dense unburned fuel in the wake of the wave, \hl{which is consumed by a surface turbulent flame.  Furthermore, consumption rates would be enhanced by Richtmyer-Meshkov instability associated with the transverse shock wave, Feature (4).} In fact, this region burns out only slightly faster than the observed experimental rate.  In the experiment, the pocket burns up completely within 7 frames (70{\gmu}s), while the simulation burns up completely within 5 frames (50{\gmu}s).  \hl{Using the method described in} \cref{sec.burningrate}, \hl{the turbulent flame speed of this experiment is found to be $\hat{S}_t/\hat{S}_L\approx6.6$ ($\hat{S}_t\approx 110\textrm{ms}^{-1}$).  This is comparable to the experimental flame speed found in the current study,} \cref{sec.experiments}, \hl{where $\hat{S}_t/\hat{S}_L\approx7.3$ ($\hat{S}_t\approx 120\textrm{ms}^{-1}$)}.  Feature (5), in figure \ref{fig.evolution2}, is a hot spot that forms behind a bifurcation of the Mach shock and is only clearly visible in the numerical simulation.  This bifurcation of the Mach stem near the triple point was also observed in the previous comparison to experiment, figure \ref{fig.evolution}.  As the Mach stem, Feature (1) continues to evolve, the combustion zone becomes further decoupled from the shock, as observed by an increased thickness of dense unburned fuel behind the shock.  Furthermore, a new pocket of dense unburned fuel forms in the wake of the triple point path, denoted by Feature (6) of figure \ref{fig.evolution2_cont}.  This pocket is initially consumed from its edges via turbulent flame.   The passage of the reflected transverse shock wave, Feature (7), further contributes to turbulent mixing through Richtmyer-Meshkov.

\begin{figure}
  \centering
  \includegraphics[scale=0.8]{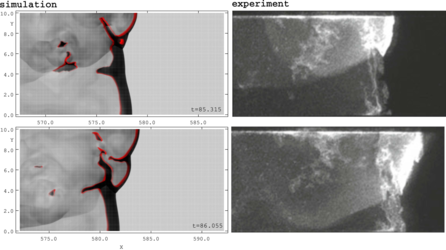}
  \caption[]{\hl{Numerical density evolution with superimposed chemical reaction rate ($\overline{\dot{\omega}}$), shown in red (left), and the corresponding experimental self-luminous images} \citep{Kiyanda2013} \hl{(right). The time in between images is $\Delta t=0.74$ (20.0\gmu s).}}
  \label{fig.reactionrate}
\end{figure}

\hl{Finally, for the $H=10$ simulation, figure} \ref{fig.reactionrate} \hl{shows how the instantaneous reaction rate obtained numerically compares with the corresponding experimental photographs of} \cite{Kiyanda2013}, \hl{which recorded the self-luminosity signal.  In this figure, the instantaneous rate of reaction, $\dot{\omega}$, is superimposed onto two consecutive density evolution plots to show where chemical reactions are occurring, and the intensity.  In both the experiment and the simulation, chemical reactions appear to be very intense behind the Mach shock, where short ignition delays are expected due to the overdriven shock.  Also, combustion of the unburned gas pockets also becomes intensified with passage of the transverse shock wave.  Finally, for the unburned pockets of gas, the chemical reactions predominantly occur on the surfaces, and not uniformly throughout.  This suggests that turbulent mixing is the dominant mechanism through which the unburned pockets are consumed. It is therefore of particular interest to quantify the turbulent mixing rates and determine their impact on burn out rates of the pockets.  Furthermore, it is of interest to determine the overall impact such burning has on the observed structure and cell pattern of the detonation wave.}

\subsection{\hl{Quantitative analysis:  Turbulent flame speeds}}
\label{sec.turbFlame}

\hl{For the detonation simulations presented in} \cref{sec.qualitycompare}, \hl{where $C_\kappa=6.7$ for both $H=10$ and $H=20$, it is of particular interest to estimate and compare the turbulent flame speed at which the pockets of unreacted gas, labelled as feature (4) in figure} \ref{fig.evolution} \hl{and feature (3) in figure} \ref{fig.evolution2}, \hl{burn up.  In contrast to the experiments, the turbulent flame speed of the pockets, in the simulations, are readily available by considering the rates at which reactant mass is consumed within the one-dimensional `samples' of the CLEM subgrid.  For the pockets, the local turbulent flame speeds are estimated from}
\begin{equation}
   S_{t,\textrm{local}}=\frac{\dot{m}}{\rho_u A_c}
   \label{eqn.flamespeed}
\end{equation}%
\hl{where $\dot{m}$ is the instantaneous rate of reactant mass consumption, $\rho_u$ is the density of the upstream unburned reactant, located to within $0.5\Delta_{1/2}$, and $A_c$ is the cross sectional area through which the flame propagates against the unburned reactant within each `sample'.  In this case, $A_c$ is known and constant for all CLEM domains.  In this approach, mass is conserved since the burning rate is balanced by the mass flow rate of unburned reactant.  Furthermore, this method was previously applied to calculate turbulent flame speeds in a stand-alone one-dimensional CLEM subgrid formulation} \citep{Maxwell2015}.  \hl{Finally, the global turbulent flame speed is obtained by ensemble averaging the flame speeds on the pocket surface in both space and time.  Thus}
\begin{equation}
   S_t=\sum_{1}^{n}S_{t,\textrm{local}}/n
   \label{eqn.flamespeed2}
\end{equation}%
\hl{where $n$ is the number of samples acquired on the pocket surface.} 

\hl{Using this method, the turbulent flame speeds of the unburned pockets are found to be $S_t/S_L=3.74$ ($\hat{S}_t=61.3\textrm{ms}^{-1}$) and $S_t/S_L=3.63$ ($\hat{S}_t=59.5\textrm{ms}^{-1}$) for the $H=20$ and $H=10$ simulations, respectively.  These values differ, by a factor of 2, from those found experimentally, which were $S_t/S_L\approx7.3$ ($\hat{S}_t=120\textrm{ms}^{-1}$) and $S_t/S_L\approx6.6$ ($\hat{S}_t\approx 110\textrm{ms}^{-1}$) for $H=20$ and $H=10$, respectively.  The difference in simulation values from the experimental ones are noted in the difficulties of evaluating the true volume and surface area solely from Schlieren images in the experiments.  The simulation and experiment, however, both predict turbulent flame speeds of roughly half an order to an order of magnitude of the turbulent flame speed.}

\begin{figure}
  \centering
  \includegraphics[scale=1.2]{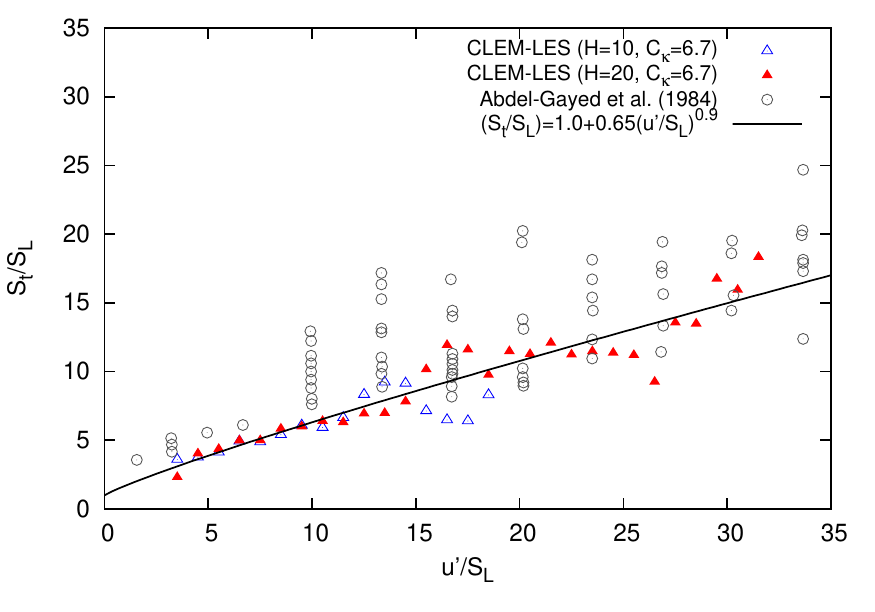}
  \caption[]{\hl{Average turbulent flame speeds ($S_t/S_L$) vs turbulent intensities ($u'/S_L$), obtained from simulation and compared to experiments of} \cite{AbdelGayed1984}.}
  \label{fig.turbulentFlameSpeeds}
\end{figure}

\hl{To gain further insight into the effect of local turbulent mixing rates on turbulent flame speeds, it is possible to obtain average turbulent flame speeds for given ranges of turbulence intensity values, $u'$, since the subgrid kinetic energy is known.  In fact, the turbulence intensity is simply}
\begin{equation}
   u'=\sqrt{\frac{2}{3}k^{sgs}}
   \label{eqn.turbIntense}
\end{equation}%
\hl{Figure} \ref{fig.turbulentFlameSpeeds} \hl{shows the average turbulent flame speeds, $S_t/S_L$, in the wake of the simulated detonations according to their turbulence intensities, $u'/S_L$.  For each value of $u'/S_L$, the flame speeds are averaged within intervals of $u'/S_L\pm0.5$.  Also shown in the figure, are fan-stirred bomb experiment measurements of} \cite{AbdelGayed1984} \hl{for stoichiometric methane-air mixtures at atmospheric pressures.  Clearly, the simulation exhibits larger average turbulent flame speeds in regions of higher turbulence intensity.  In fact, much like the experiments of} \cite{AbdelGayed1984}, \hl{$S_t/S_L$ appears to increase proportionately with $u'/S_L$.  Turbulence intensities were not recorded above the values shown in the figure, as such high velocity fluctuations have not had sufficient time to develop on the pocket surfaces.  For the turbulence intensities observed, the results are in good agreement with the experiments.}

\subsection{Quantitative analysis:  The reaction zone and hydrodynamic thickness}

{In order to determine the effect that turbulent mixing intensity has on the reaction zone thickness, the average thickness of the structure ($\Delta_R$) has been measured for various $C_\kappa$ values for the domain height of $H=20$.  This was achieved by following the procedure to generate the Favre-averaged reactant profiles in figure \ref{fig:ck_profiles}.  {Figure} \ref{fig:ck_thickness} {thus shows the reaction zone thickness ($\Delta_R$), defined here as the distance from the shock wave, $x=x_s$, to where $\tilde{Y}(x)<2\%$, as a function of $C_\kappa$.} Clearly, there is an increasing trend in thickness with $C_\kappa$.  In fact, the thickness appears to increase linearly with $C_\kappa$ for the range of values simulated here.  A line of best fit (obtained from linear regression of each data point), $\Delta_R=1.19C_\kappa+4.49$, is also indicated in the figure.  Also shown in the figure are error bars indicating the standard deviation of thickness for each value of $C_\kappa$.

\begin{figure}
  \centering
  \includegraphics[scale=1.2]{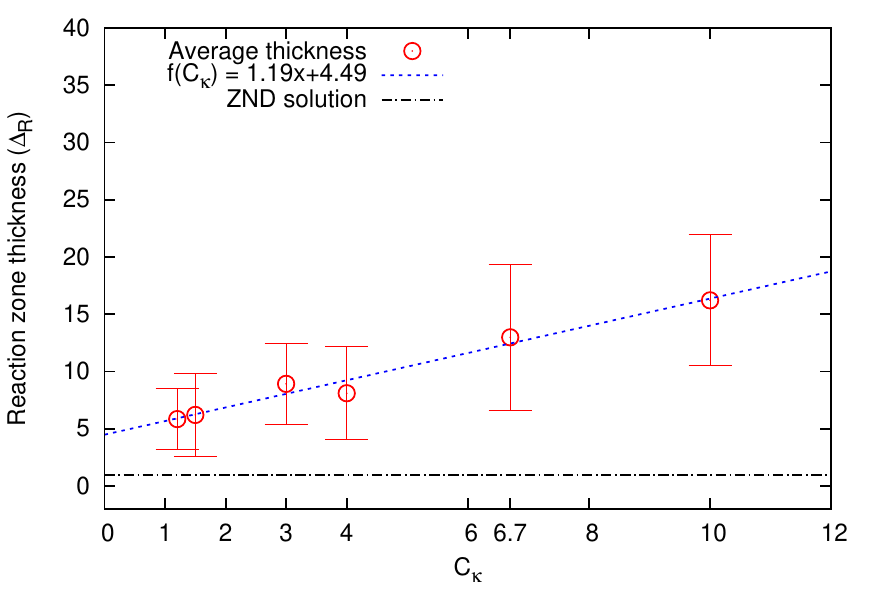}
  \caption[Effect of turbulence intensity (through $C_\kappa$) on reaction zone thickness ($\Delta_R$) for the CLEM-LES.]{Effect of turbulence intensity (through $C_\kappa$) on the mean reaction zone thickness ($\Delta_R$) for the CLEM-LES (all with resolution $b=1/32$ and $N=16$ elements per LES cell).   The error bars indicate the standard deviation in $\Delta_R$ for each simulation.  Note: $\Delta_R$ is normalized by $\hat{\Delta}_{1/2}$.}
  \label{fig:ck_thickness}
\end{figure}

{To quantitatively compare the CLEM-LES to experimental observations in terms of the reaction zone thickness, or distance it takes for the pockets of unburned gas to burn, results from section} \ref{sec.CK_effect} {are considered.  In figure} \ref{fig:ck_thickness}, {it was observed that the average reaction zone thickness captured by all values of $C_\kappa$, for $H=20$, was in the range $5<\Delta_R<17$.  More specifically, for $C_\kappa=6.7$ and $C_\kappa=10.0$, the average thickness was found to be $\Delta_R=13.0$ and $\Delta_R=16.2$, respectively.  These computed average thicknesses much higher than previous reports of burn out distances between} $\sim4-6\hat{\Delta}_{1/2}$ \citep{Radulescu2007b}, which were measured from Euler simulations with significant numerical viscosity. Upon close inspection, however, of experimental images shown in \cref{sec.experiments}, {burn out distances were estimated as high as} $\sim9\hat{\Delta}_{1/2}$, {as can be seen in figure} \ref{fig:Bhattacharjee_thickness}. {Thus, compared to the experiment in figure \ref{fig:Bhattacharjee_thickness}, $C_\kappa=6.7$ yields an error of $\sim44\%$, while $C_\kappa=10.0$ yields an error of $\sim80\%$.  Despite these large errors, however, it should be noted that the simulations for $C_\kappa=6.7$ and $C_\kappa=10.0$ have standard deviations of $\Delta_R\pm6.4$ and $\Delta_R\pm5.7$, respectively.  Thus, a large range of possible expected thicknesses lie anywhere from $6.6<\Delta_R<22$.  Furthermore, the current experiments, which are stochastic in nature, have not quantified the probability and variance of the reaction zone thickness possible during propagation .}

\begin{figure}
  \centering
  \includegraphics[scale=0.8]{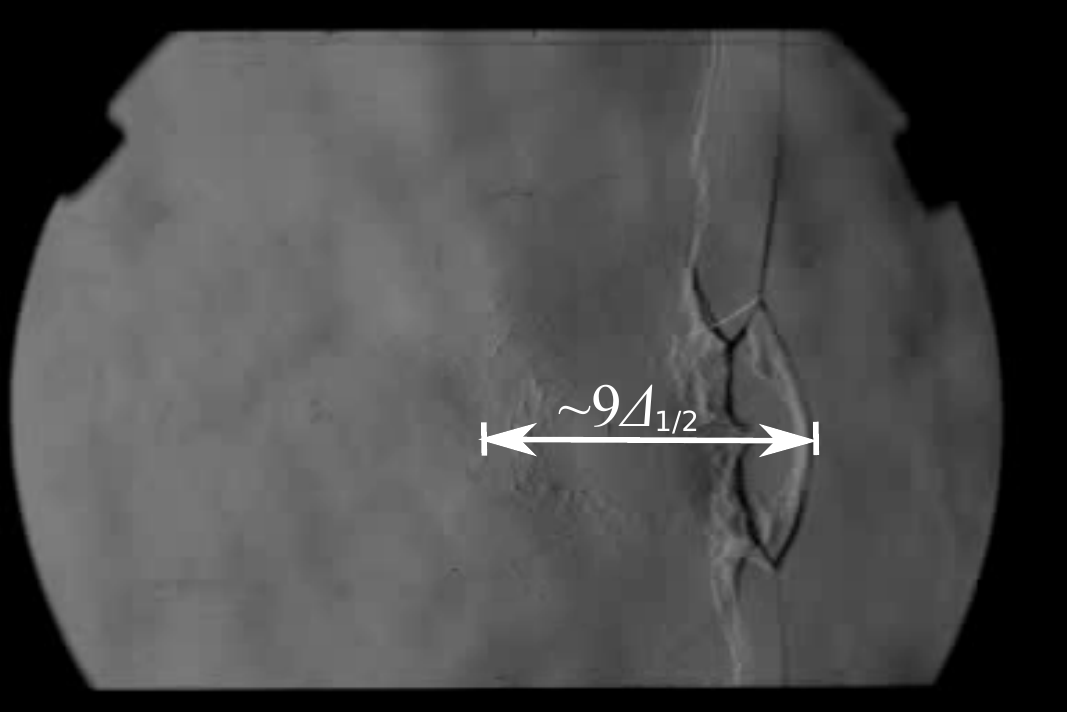}
  \caption[Schlieren photograph which shows that pockets of unburned gas can take up to $\sim9\hat{\Delta}_{1/2}$ behind the leading shock to burn up.]{Schlieren photograph (this study), {which shows that pockets of unburned gas can take up to} $\sim9\hat{\Delta}_{1/2}$ {behind the leading shock to burn up.}}
  \label{fig:Bhattacharjee_thickness}
\end{figure}

Finally, the reaction zone thickness has been well correlated with the detonation cell size, where chemical reactions are completed roughly within one-cell cycle \citep{Lee1984}.  From figure \ref{fig:ck_openshutters}, the cell size $\lambda\approx10$ when $C_\kappa=6.7$.  This compares well with the average reaction zone thickness, $\Delta_R=13.0$.  Furthermore, the entire hydrodynamic thickness of the wave ($\Delta_H$), or distance from the leading shock to the trailing sonic plane (or CJ state), is also well correlated with the cell size: $\Delta_H=6.5\lambda$ \citep{Lee1984}.  In figure \ref{fig:av_pg}, the ensemble space and time averaged pressure profile for $\bar{p}(x)$ is presented for the $C_\kappa=6.7$ and $H=20$ case and compared to the ZND solution for $p(x)$.  For this case, it takes approximately $\sim50\Delta_{1/2}$ for the pressure to reach the CJ-solution state, which corresponds to the sonic plane in the wake of the wave.  This is consistent with findings in \cite{Lee2005} for highly unstable detonations involving methane at similar conditions.  Furthermore, since $\lambda\approx10$ for $C_\kappa=6.7$, then $\Delta_H\approx5\lambda$.  This result is in the same order as that \hl{reported} by \cite{Lee1984} \hl{from previous experimental correlations}.  To summarize the findings here, increasing the value of $C_\kappa$ was found to generate a larger reaction zone thickness.  This, in turn, is believed to lead to an increased hydrodynamic thickness and cell size, due to their dependence on the reaction zone thickness.

\begin{figure}
  \centering
  \includegraphics[scale=1.4]{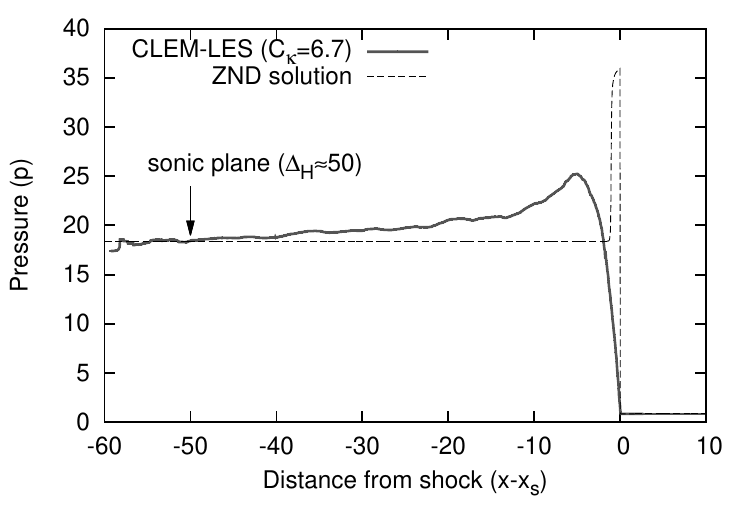}
  \caption[Ensemble space and time averaged pressure profile ($\bar{p}(x)$) for the $C_\kappa=6.7$ and $H=20$ simulation and compared to the ZND solution for $p(x)$.]{Ensemble space and time averaged pressure profile ($\bar{p}(x)$) for the $C_\kappa=6.7$ and $H=20$ simulation and compared to the ZND solution for $p(x)$. Note: Distances $x$ and $x_s$ are normalized by $\hat{\Delta}_{1/2}$.  The pressure $p$ is normalized by $\gamma p_o$.}
  \label{fig:av_pg}
\end{figure}

\subsection{Qualitative comparison of cell patterns with experiments}

To further compare the CLEM-LES with experiment, in terms of cell patterns and irregularity, numerically obtained soot foils for both $H=10$ and $H=20$ at $C_\kappa=6.7$ was compared to those obtained experimentally for $\textrm{CH}_4+\textrm{2O}_2$ at $\hat{p}_o=3.5$kPa, courtesy of \cite{Radulescu2005}, and $\textrm{CH}_4+\textrm{2O}_2+0.2$Air at $\hat{p}_o=11$kPa \citep{Austin2003}.  {For $H=20$, the numerical soot foil, presented in figure} \ref{fig:Austin2003_ch4soot}, {corresponds exactly to the soot foil image, of figure} \ref{fig:ck_openshutters}, {for $C_\kappa=6.7$.  Although the experimental soot foil, also shown in figure} \ref{fig:Austin2003_ch4soot}, {was obtained for a smaller channel height (127mm) and also for a much higher pressure, the two are comparable in terms of scaling through the half-reaction length.  For the experiment} \citep{Austin2003}, $\hat{\Delta}_{1/2}=5.1$mm.  {This value was obtained in this study using the Cantera libraries} \citep{Goodwin2013} {and the GRI-3.0 detailed kinetic mechanism} \citep{Smith2013}, {consistent with the procedure in} \cite{Maxwell2016}.  {Thus, the experimental soot foil was estimated to be} $\sim25\hat{\Delta}_{1/2}$ {high.  This is comparable to the simulation, whose domain is} $20\hat{\Delta}_{1/2}$ {in height.  In the portion of the numerical soot foil presented in the figure, the overall cell size and pattern matches well to the experiment.  In the experiment, however, a very distinct substructure was observed.  This was observed by the presence of much smaller cells within the larger and more predominant cell structure.  Although such a fine substructure is not so obvious in the simulation, there are still some streaks on the numerical soot foil, which are indicative of triple point paths associated with cell bifurcations.  The lack of detailed substructure in the simulation could be an artifact of the supergrid resolution,  differences in mixture properties, or most likely associated  with the limitation of the one-step model in capturing the very short duration energy release of methane combustion and the associated instabilities that exists on that finer small scale.  Furthermore, it is likely that at the resolution presented here, 32 grids per $\hat{\Delta}_{1/2}$, has effectively filtered out the fine details associated with the expected substructure.  It is noted that higher resolution simulations were not conducted for $C_\kappa=6.7$.  Despite this lack of fine-scale detail for the substructure, the overall cell behaviour and irregularity is captured well by the simulation.}

\begin{figure}
  \begin{subfigure}[b]{0.55\textwidth}
  \raggedright Simulation:\\
  {\includegraphics[scale=0.73125]{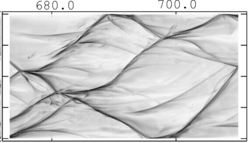}}
  \end{subfigure}
  \begin{subfigure}[b]{0.5\textwidth}
  \raggedright Experiment:\\
  \vspace{0.2cm}{\includegraphics[scale=0.5]{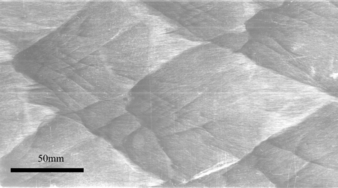}}\vspace{0.1cm}
  \end{subfigure}
  \caption[Numerical soot foil obtained for the CLEM-LES with $C_\kappa=6.7$ and $H=20$ compared with an experimental soot foil for $\textrm{CH}_4+\textrm{2O}_2+0.2$Air at $\hat{p}_o=11kPa$ \citep{Austin2003}.]{{Numerical soot foil obtained for the CLEM-LES with $C_\kappa=6.7$ and $H=20$ (left) compared with an experimental soot foil for $\textrm{CH}_4+\textrm{2O}_2+0.2$Air at} $\hat{p}_o=11kPa$ \citep{Austin2003} {(right).  Note: the channel height for the experiment is} $\sim25\hat{\Delta}_{1/2}$ \hl{(127mm), while the simulation is $20\hat{\Delta}_{1/2}$ (203mm).}}
  \label{fig:Austin2003_ch4soot}
\end{figure}

\begin{sidewaysfigure}
  \vspace{14.0cm}
  \raggedright Experiment:\\
  \centerline{\includegraphics[scale=1.0]{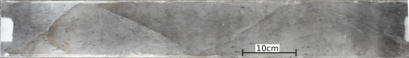}}
  \raggedright Simulation:\\
  \centerline{\includegraphics[scale=1.0]{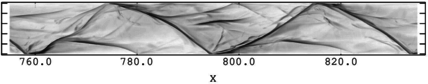}}  
  \raggedright Experiment:\\
  \centerline{\includegraphics[scale=1.0]{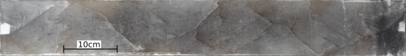}}
  \raggedright Simulation:\\
  \centerline{\includegraphics[scale=1.0]{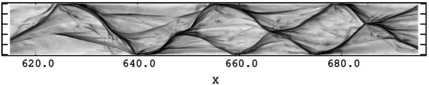}}  
  \caption[{Numerical soot foils obtained for the CLEM-LES with $C_\kappa=6.7$ and $H=10$ compared with two experimental soot foil for $\textrm{CH}_4+\textrm{2O}_2$ at} $\hat{p}_o=3.5kPa$.]{{Numerical soot foils obtained for the CLEM-LES with $C_\kappa=6.7$ and $H=10$ compared with two experimental soot foil for $\textrm{CH}_4+\textrm{2O}_2$ at} $\hat{p}_o=3.5kPa$, {courtesy of M. Radulescu.  Note: the channel height for the experiment is also} $10\hat{\Delta}_{1/2}$.}
  \label{fig:Radulescu2015_ch4soot}
\end{sidewaysfigure}

{For $H=10$, figure} \ref{fig:Radulescu2015_ch4soot} shows numerical soot foils obtained for two different portions of the numerical domain and compares them to soot foils of experiments conducted in \cite{Radulescu2005}.  {Here, the experimental soot foils were obtained using the same experimental apparatus described in} \cite{Kiyanda2013}.  {For both experimental soot foils, the numerical simulation captures well the cell size behaviour and irregularity.  In general the channel height allows only 1/2 cell to form.  The cells, however, occasionally bifurcate, giving rise to the formation of cells on the order of the channel height.  This behaviour can be observed in the bottom frames of figure} \ref{fig:Radulescu2015_ch4soot}.  {Finally, it is noted that a prominent substructure was not observed experimentally or numerically for $H=10$, as was observed experimentally in figure} \ref{fig:Austin2003_ch4soot}.  {This could be attributed to differences in the mixture or conditions (i.e., the pressure) at which tests were performed.}

\subsection{Quantitative analysis:  Velocity of the wave}
\label{sec.pdf_velocity}

{To gain insight on why increasing $C_\kappa$ has the effect of generating larger hydrodynamic structures,} the subgrid kinetic energy associated with random velocity fluctuations ($k^{sgs}$) is Favre-and ensemble averaged for each $C_\kappa$ case using the same procedure as was done for the reactant profiles, i.e., equations \eqref{eqn.Favre_space}-\eqref{eqn.ensemble_time}.  The averaged profiles obtained for $k^{sgs}$ at various values of $C_\kappa$ are thus shown in figure \ref{fig:ke_vSck}.  The principle observation made here is that as $C_\kappa$ is increased, more subgrid kinetic energy is generated.  This is not surprising since $k^{sgs}$ is a function $C_\kappa$, as demonstrated in \cite{Maxwell2016}. This increase in $k^{sgs}$ with $C_\kappa$ thus generates more frequent stirring events on the CLEM subgrid\hl{, and therefore faster burning rates on pocket surfaces. Despite this faster mixing and burning, larger pockets of unburned gas are able to form in the wake.  This suggests an overall velocity deficit develops on the wave front, allowing for increased ignition delays.}

\begin{figure}
  \centering
  \includegraphics[scale=1.2]{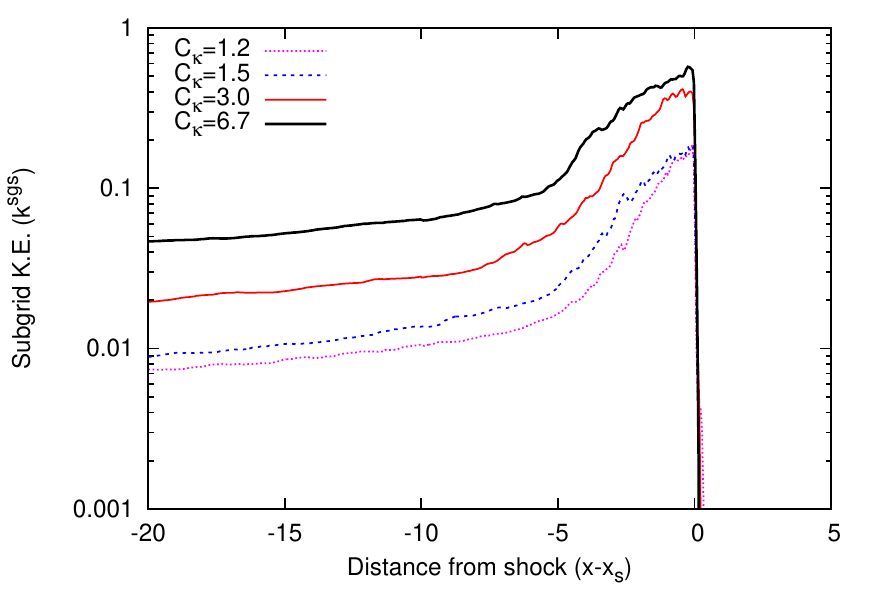}
  \caption[Effect of $C_\kappa$ on the averaged subgrid kinetic energy  ($k^{sgs}$) profile for the CLEM-LES.]{Effect of $C_\kappa$ on the averaged subgrid kinetic energy  ($k^{sgs}$) profile for the CLEM-LES. Note: Distances $x$ and $x_s$ are normalized by $\hat{\Delta}_{1/2}$.}
  \label{fig:ke_vSck}
\end{figure}

To quantify the statistical distribution of velocities experienced by the wave front, and such a distribution was affected by $C_\kappa$, probability density functions (PDFs) were constructed from the numerical simulations and shown in figures \ref{fig:speed_pdf_h20} and \ref{fig:speed_pdf_h10} for $H=20$ and $H=10$, respectively.  Also shown in these two figures, are the experimental PDFs, previously shown in figure \ref{fig:exp_pdf}.  To construct the PDFs from the numerical simulations, velocity measurements were taken on the top and bottom walls for a total of 500 data points in each simulation.  The PDFs were then evaluated at $D\pm0.5$ intervals and normalized by $D_{avg}$ accordingly.  For the CLEM-LES, conducted at various $C_\kappa$ values, and also an Euler simulation at the same supergrid resolution ($b=1/32$), the average propagation speed is always within 1\% error of the CJ-value, where $D_{CJ}=6.30$ \hl{(2240$\textrm{ms}^{-1}$)}.

\begin{figure}
  \centering
  \includegraphics[scale=1.2]{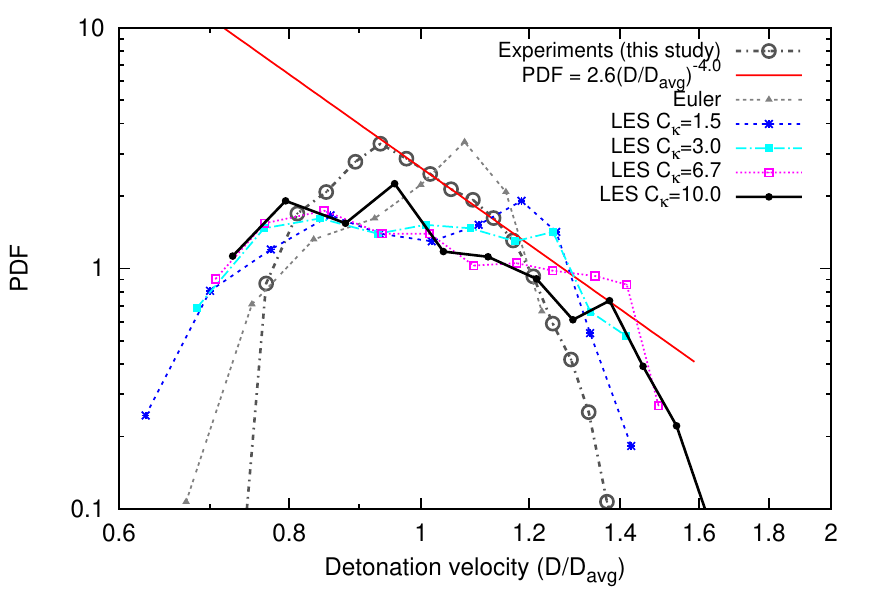}
  \caption[PDF of a detonation wave, in a channel with $H=20$, having a certain velocity ($D$) at any given moment and location.]{{PDF of a detonation wave, in a channel with $H=20$, having a certain velocity ($D/D_{avg}$) at any given moment and location.  Also shown is a PDF compiled from experiments (this study)}. \hl{Note:  $D_{avg}=5.19$ (1850$\textrm{ms}^{-1}$) for the experiment and $D_{avg}=6.30$ (2240$\textrm{ms}^{-1}$) for the simulations.}}
  \label{fig:speed_pdf_h20}
\end{figure}

\begin{figure}
  \centering
  \includegraphics[scale=1.2]{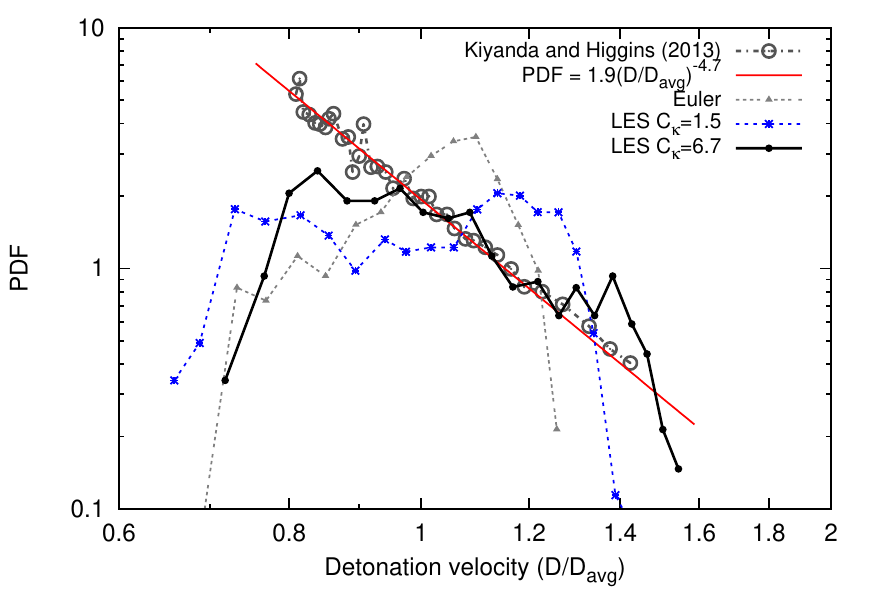}
  \caption[PDF of a detonation wave, in a channel with $H=10$, having a certain velocity ($D$) at any given moment and location.]{{PDF of a detonation wave, in a channel with $H=10$, having a certain velocity ($D/D_{avg}$) at any given moment and location.  Also shown is a PDF compiled from} \cite{Kiyanda2013}. \hl{Note:  $D_{avg}=5.53$ (1970$\textrm{ms}^{-1}$) for the experiment and $D_{avg}=6.30$ (2240$\textrm{ms}^{-1}$) for the simulations.}}
  \label{fig:speed_pdf_h10}
\end{figure}

Clearly, from figures \ref{fig:speed_pdf_h20} and \ref{fig:speed_pdf_h10}, as $C_\kappa$ increases, the likelihood, or probability, of the detonation wave to travel at speeds below the average CJ value also increases.  This is especially true for $C_\kappa\ge6.7$.  At these high values of $C_\kappa$, the detonation tends to favour a greater chance of having wave speeds below the average propagation speed value, i.e., when $(D/D_{avg})<1$.  Conversely, at lower values of $C_\kappa$, and the Euler simulation, velocities above the average propagation speed are favoured.  This would lead to much shorter ignition delays, thus explaining the relatively quick burning times associated with the observed shortened reaction zone thickness, previously shown in figure \ref{fig:ck_thickness}.  For higher $C_\kappa$ values, since the wave spends more time at below CJ-velocities, most of the unburned gas that is shocked by the wave front has a much longer ignition delay compared to the ZND model.  For this reason, unburned pockets of gas are able to form in the wake, which eventually burn up through turbulent mixing.  This favouring of velocities below the average propagation speed thus lengthens the overall hydrodynamic structure of the wave.  Consequently, the cells also increase in size and become more irregular in appearance, as observed in the cell patterns of figure \ref{fig:ck_openshutters}.  {Since higher $C_\kappa$ values inherently generate more random fluctuations, the cell irregularity would also be expected to increase.

In general, the PDFs collected for $C_\kappa\ge6.7$ compared well to experiments.  In fact, simulations for $C_\kappa\ge6.7$ reproduce well the decaying behaviour of the wave velocity, where the probability of the wave speed exhibits power-law dependence on wave speeds above the favoured value.  For $H=20$, although the numerical simulations for $C_\kappa\ge6.7$ do not collapse onto this correlation, the same decaying trend is observed.  {For $H=20$, the most probable wave speed favoured, experimentally}, is $(D/D_{avg})=0.93\pm0.04$ {with a peak PDF value of 3.3. For the simulation at $C_\kappa=6.7$, the most probable velocity was $(D/D_{avg})=0.96\pm0.04$ with a peak PDF value of 1.74.  This favoured velocity value has $8.6\%$ error compared to experiment.  For $C_\kappa=10.0$, the most probable velocity was $(D/D_{avg})=0.85\pm0.04$ with a peak PDF value of 2.25.  This agrees well for the favoured velocity value, with only $3.2\%$ error.  Despite this, however, the range of expected velocities on the wave front is much larger in the simulations compared to experiment.  These differences in peak PDF values and expected ranges of velocities are believed to be influenced by several factors.  First, the numerical simulations do not account for energy losses which are present in the experiments.  Therefore, the detonation, in the simulations, has higher velocities.  Furthermore, cell patterns for a channel height of $H=20$ are very stochastic and irregular.  The experimental PDF in figure \ref{fig:speed_pdf_h20} was compiled for only 4 different experiments.  It is likely that much more experiments are required in order to capture the events in the tails of the PDF, beyond the velocity limits currently obtained.  Finally, velocity measurements are captured experimentally every 11.53{\gmu s}, which may filter out any high speed and short-lived velocity fluctuations, which may occur beyond the current PDF limits.  It is likely that in reality, the experimental PDF should have a larger range with a smaller peak PDF value.

For $H=10$, much better agreement to experiment is observed when $C_\kappa=6.7$.  The most probable wave speed favoured, experimentally} \citep{Kiyanda2013}, {is $(D/D_{avg})=0.814\pm0.005$ with a peak PDF value of 6.15. In this case, the simulation at $C_\kappa=6.7$ has a favoured velocity of $(D/D_{avg})=0.84\pm0.04$ with a peak PDF value of 2.55.  This also agrees well for the favoured velocity value, with only $1.0\%$ error.  This better agreement is likely attributed to the fact that a domain height of $H=10$ effectively mode-locks the detonation in such a way that much less variation in cell size was observed.  This was seen in the soot foils presented in figure \ref{fig:Radulescu2015_ch4soot}. Clearly, the analysis conducted here supports the need to calibrate $C_\kappa$ in order to ensure the correct velocity probability trend, and qualitative features, are obtained numerically.}

\subsection{\hl{3D effects and the validity of the two-dimensional LES approach}}
\label{sec.3D_effect}

\hl{Finally, in order to further validate the two-dimensional numerical results obtained throughout this work, a thin channel 3D simulation has been conducted for the $H=10$ case, with $C_\kappa=6.7$ and a domain width of 2$\hat{\Delta}_{1/2}$.  This corresponds roughly to the experimental setup of} \citep{Kiyanda2013}.  \hl{The grid resolution used for the 3D simulation was $b=1/16$ with $N=16$ subgrid elements within each cell.  Figure} \ref{fig:3dsootfoil_h10} \hl{shows a portion of soot foils which were obtained on the side and top walls of the domain.  More specifically, these are obtained at $z=0$ in the $x-y$ plane and $y=10$ in the $x-z$ plane, respectively.  Here, $x$ is the direction along the channel length, $y$ is along the channel height, and $z$ is along the channel width. Upon comparing the side wall soot foil with that of the 2D LES simulation in Figure} \ref{fig:Radulescu2015_ch4soot}, \hl{the cell size obtained is comparable.  In both the 2D and 3D simulations, the resulting cell size is one-half cell per channel height throughout most of the domain.  Furthermore, the soot foil obtained in the 3D simulation does not exhibit a visible sub-structure, as was the case for the 2D simulation in figure} \ref{fig:Radulescu2015_ch4soot}. \hl{Finally, the vorticity streaks on the top wall soot foil suggest the resulting cell pattern is predominantly two-dimensional for the thin-channel.}

\hl{To further investigate the resulting flow pattern of the 3D simulation, various sections of density profiles are extracted in the $x-y$, $z-y$, and $x-z$ planes for a single instance in time and presented in figure} \ref{fig:3dsections_h10}.  \hl{This was done for an instance in time just before a triple point reflects on the top wall, where a pocket of unburned gas has nearly formed in the wake of the wave. In three sections taken in the $x-y$ plane, at $z=0.25$, $z=1.0$, and $z=1.75$, it is clear that the dense pocket of unreacted gas does not have a uniform shape along the width of the channel. Upon investigating the sections along various $x$ locations in the $z-y$ plane and $y$ locations in the $x-z$ plane, it is clear that the burning of the pocket of unreacted gas indeed occurs in all three-dimensions.  The presence of unburned gas, however, varies much more significantly in the $x$ and $y$ directions compared to the channel width, in the $z$ direction.  Significant burn-up along the channel width is only observed in frames where the remaining pocket is nearly consumed, sufficiently far from the leading shock.  This can be seen in frames in the $z-y$ plane for $x\leq734.5$ and the $x-z$ plane for $y\leq5.5$.  Finally, despite the non-uniformities in the pocket shape observed across the channel width, and the three-dimensional structure of the pocket burn-out, the position of the incident and Mach shocks remain uniform throughout the channel width.}

\begin{figure}
  \centering
  \includegraphics[scale=0.5]{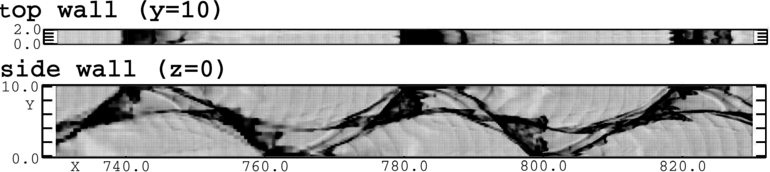}
  \caption[3D sootfoils.]{\hl{Numerical soot foils obtained for the 3D CLEM-LES for $C_\kappa=6.7$ and $H=10$.}}
  \label{fig:3dsootfoil_h10}
\end{figure}

\begin{figure}
  \centering
  \includegraphics[scale=0.8]{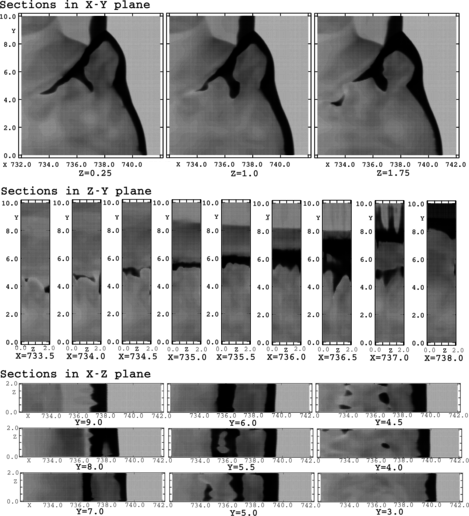}
  \caption[3D sections]{\hl{Profile sections of density extracted in the $x-y$, $z-y$, and $x-z$ planes for the 3D CLEM-LES.}}
  \label{fig:3dsections_h10}
\end{figure}

\hl{To further compare the two- and three-dimensional simulations, in terms of the rate at which the unreacted gas is consumed, the Favre-averaged reactant profiles are presented in Figure} \ref{fig:ck_profiles_3d}.  \hl{In this figure, it is clear that the reactant profiles of the three-dimensional simulation match closely the profiles obtained from the two-dimensional simulation, for $C_\kappa=6.7$.   In fact, both the 2D and 3D simulations have an average hydrodynamic thickness of $\Delta_R=8.2$ with standard deviations of $\Delta_R\pm2.7$ and $\Delta_R\pm2.0$, respectively.  To further compare two- and three-dimensional simulation results, the PDFs of wave propagation velocity for both simulations are presented in figure} \ref{fig:pdf_3d}, \hl{and compared to the experiment} \citep{Kiyanda2013}.  \hl{In this figure, it is clear that the two- and three-dimensional simulations both exhibit a similar PDF trend. The three-dimensional simulation however favours a slightly higher velocity, $(D/D_{avg})=0.87\pm0.04$ with a peak PDF value of 3.30 compared with the two-dimension simulation which favoured a velocity of $(D/D_{avg})=0.84\pm0.04$ and a peak PDF value of 2.55.  The two results, however, are within measurement error of each other.  The results presented in these two figures,} \ref{fig:ck_profiles_3d} and \ref{fig:pdf_3d}, \hl{suggest that despite the three-dimensional nature of local burning on unburned pocket surfaces, the use of two-dimensional LES is justified for flows with high aspect ratios, where cell sizes are sufficiently large in height compared to their channel width.  This can be attributed to the large-scale flow patterns, which are predominantly two-dimensional.  Furthermore, the use of two-dimensional LES, in this study, is further justified by the fact that the LEM subgrid accounts for the small-scale three-dimensional flow instabilities, on the unburned gas pocket surfaces, through the stirring terms, $\dot{F_T}$ and $\dot{F_Y}$, in equations} \ref{eqn.lemEnergy} and \ref{eqn.lemReactant} \citep{Kerstein1991b}.

\begin{figure}
  \centering
  \includegraphics[scale=1.2]{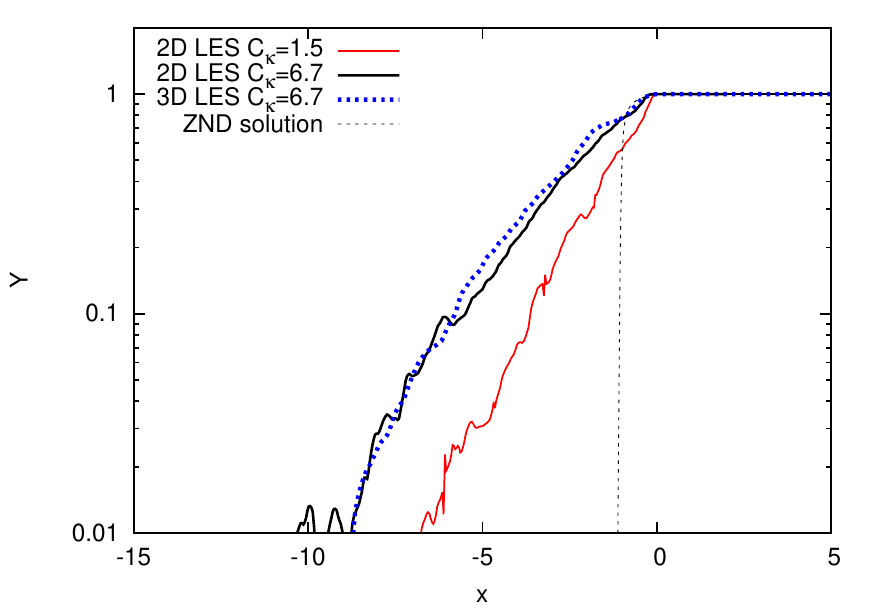}
  \caption[Effect of turbulent intensity (through $C_\kappa$) on average reactant profiles for the CLEM-LES.]{\hl{Average reactant profiles for both 2D and 3D CLEM-LES.  Note: Distances $x$ and $x_s$ are normalized by $\hat{\Delta}_{1/2}$.}}
  \label{fig:ck_profiles_3d}
\end{figure}

\begin{figure}
  \centering
  \includegraphics[scale=1.2]{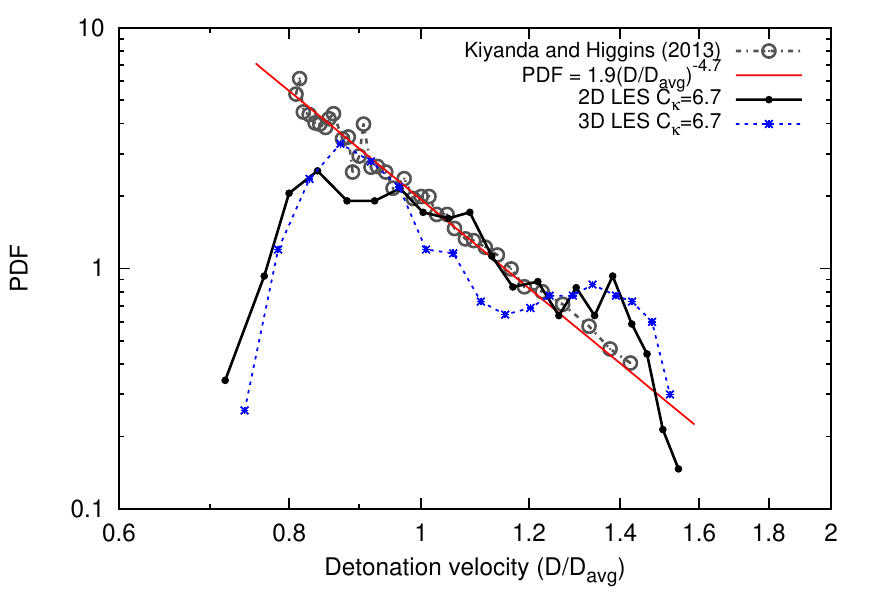}
  \caption[PDFs of a detonation wave, in a channel with $H=10$, having a certain velocity ($D$) at any given moment and location.]{\hl{PDFs of a detonation wave having a certain velocity ($D/D_{avg}$) at any given moment and location.  These PDFs have been compiled from both 2D and 3D simulations where $C_\kappa=6.7$ and $H=10$.  Also shown is a PDF compiled from} \cite{Kiyanda2013}. \hl{Note:  $D_{avg}=5.53$ (1969.1$\textrm{ms}^{-1}$) for the experiment and $D_{avg}=6.30\pm0.05$ (2243.3$\pm$17.8$\textrm{ms}^{-1}$) for the simulations.}}
  \label{fig:pdf_3d}
\end{figure}

\subsection{$C_\kappa$ as a tuning parameter}
\label{sec.tuning}

From {the numerical soot foils presented} figure \ref{fig:ck_openshutters} it is clear that the $C_\kappa$ parameter requires tuning in order to match the desired cell size and behaviour observed in the experiments of this study and \cite{Kiyanda2013}.  {From a quantitative perspective, figures} \ref{fig:speed_pdf_h20} and \ref{fig:speed_pdf_h10} {have shown that numerically obtained probability density functions of detonation velocity distributions matched well the trends obtained through experiments when $C_\kappa\ge6.7$.  In general, higher values of $C_\kappa$ allowed below average velocities to be favoured along the wave front, as was the case experimentally.  As a result, it is clear that $C_\kappa$ requires tuning in order to ensure the correct velocity distribution, and consequently the correct expected reaction zone thickness.  Furthermore,} as observed in figures \ref{fig.evolution} to \ref{fig:Radulescu2015_ch4soot}, a value of $C_\kappa=6.7$ was found adequate to form qualitative cells and burning patterns comparable to experiments.  For smaller values of $C_\kappa$, only much smaller cells appear. For larger values of $C_\kappa$, the cells increase in size until they are mode locked by the channel height.  It should be noted, however, that specific `matching' of $C_\kappa$ values can prove difficult due to the stochastic nature of both the experiments and the numerical simulations.  For this, better converged statistics would be required.  It should be further noted that even though the value of $C_\kappa=6.7$ was found to give good agreement with experimental observation, $C_\kappa=1.4-2.0$ is a more universally accepted value \citep{Bradley1992,Chasnov1991}.  This difference may be due to several reasons.  First of all, low values of $C_\kappa$ correspond to Kolmogorov's theory of the turbulence cascade in the incompressible limit \citep{Frisch2000}.  The detonation phenomenon under investigation here, on the other hand, is highly compressible.  It is possible that highly compressible regimes generate much more turbulent fluctuations than incompressible theory predicts.  This is especially true for Richtmyer-Meshkov instability where turbulent motion is generated from shock waves interacting with material interfaces.  Second of all, the \hl{large scale motions of the} experimental flow fields are essentially two-dimensional due to the thin channel cross section widths.  In fact, the aspect ratios of the channel heights to widths ($H/W$) in the experiments are $\sim10$ (this study) and $\sim4$ \citep{Kiyanda2013}.  In both cases, the channel widths are much smaller than the observed cell sizes.  Thus, straining and decay of turbulent motions along the cross sectional width may not be as prominent as would be expected in a full scale three-dimensional experiment.  Furthermore, turbulence in two-dimensions has been shown, theoretically, to exhibit back-scatter \citep{Kraichnan1967,Leith1968,Batchelor1969}, and hence generate more turbulent motions.  This has also been assumed to be the case for atmospheric flows, which are also considered essentially two-dimensional at planetary scales \citep{Lilly1966}.  In this sense, small scale turbulence is able to feed into, and amplify, larger scale turbulent motions.  \hl{Although three-dimensional burning does occur in the wake of the detonation wave at smaller scales, a three-dimensional simulation conducted in the previous section suggested that the rate of fuel burn-up is largely governed by the large-scale fluid motions along the the channel length and height.}  Since the experiments of this study and \cite{Kiyanda2013} are quasi-two-dimensional, it is possible there are more turbulent motions behind the detonation front than would be expected in a naturally occurring three-dimensional detonation wave propagation.  In this sense, it is possible the corresponding two-dimensional simulation would also require a higher $C_\kappa$ value to artificially generate more turbulence.  Furthermore, \cite{Kraichnan1971} predicts a theoretical value of $C_\kappa=6.69$ for two-dimensional turbulence, compared to 1.4 for three-dimensions. Another review of two-dimensional turbulence \citep{Danilov2000} indicates a $C_\kappa$ value of 5.8 to 7 should be adopted due to the increased turbulence associated with back-scatter. It should be noted, however, that in the current work this agreement is simply treated as a coincidence and has not been investigated further.  Other sources of discrepancy may arise from errors associated with limitations of the chemical model used, or perhaps discrepancies associated with experimental losses \citep{Maxwell2016}.  Finally, in terms of calibration, it should be noted that it is possible to implement a dynamic procedure to obtain the $C_\kappa$ value, i.e., the Localized Dynamic Kinetic Energy Model (LDKM) method \citep{Schumann1975,Menon1996,Menon2011}.  Although this is possible, validation and verification would still be required.  Furthermore, the method may be more appropriate for three-dimensional simulations and would therefore increase computational expense significantly.  Also, the dynamic procedure may not adequately account for increased turbulent fluctuations due to compressibility effects.  For now, the static method of prescribing $C_\kappa$ proves advantageous in two-dimensions for examining the effect of turbulence intensity on detonation propagation patterns and behaviours.


\section{Conclusions}
\label{sec.conclusion}
In this work, experiments and numerical simulations, involving methane-oxygen, have been conducted to identify the mechanisms which contribute to irregular detonation propagation.  Of particular interest was to determine the mechanisms leading to the formation of unburned pockets of reactive fuel in the wake of an irregular detonation, and how the burning of these pockets influence the overall cell pattern, size, and structure during propagation.   \hl{Furthermore, the burning rate on the surface of such pockets was determined through both experiment and simulation to be on the order of $\sim60-120\textrm{ms}^{-1}$ ($S_t/S_L\approx3.7-7.3$), and was found to vary proportionally to the local turbulence intensity.}  To this effect, experiments provided a general qualitative insight into the role of turbulent mixing on the propagation characteristics, as well as the necessary data, qualitatively and quantitatively, to validate the numerical strategy, from which a detailed investigation was carried out.

From experiments, the Richtmyer-Meshkov flow instability, behind triple point paths, plays a major role in the burning of such pockets. On the other hand, Kelvin-Helmholtz is believed to contribute only a minor role in such burning.  In fact, KH instability growth rates, in the presence of burned and unburned fuel interfaces along slip-lines behind triple-points, can actually be damped by the decoupling of velocity and thermal gradients in high activation energy mixtures \citep{Massa2007}.  In general, flow instabilities were found to generate turbulent mixing behind the front of the wave, thus influencing the overall burning rate of the pockets.  Furthermore, turbulent motions were found to entrain burned products into regions of shocked, and unburned gas, also contributing to the overall burning rate of unburned fuel in the wake. In order to investigate, in detail, how the turbulent mixing rates of these pockets affect the overall propagation characteristics of the wave, in terms of cell pattern and size, irregularity, and reaction zone thickness, numerical simulations have been carried out.  In order to resolve the small scale mixing and chemical reactions, which are often very challenging to resolve for flows which are highly compressible, reactive, and contain rapid transients in pressure and energy, the Compressible Linear Eddy Model for Large Eddy Simulation (CLEM-LES) is applied.  The strategy itself serves as extended work to a previously developed Linear Eddy Model formulation for Large Eddy Simulation \citep{Menon2011,Menon1996,Chakravarthy2000,Sankaran2005} whose capabilities were limited to modelling only weakly compressible flows in multiple dimensions.  To this effect, the CLEM-LES strategy was validated against experiments accordingly.  The major contribution of the current work, in terms of advancing numerical strategies, was to provide adequate closure of the reaction rate term, $\overline{\dot{\omega}}$, in equation \eqref{eqn.LESenergy2} for modelling and understanding detonation behaviour.  In fact, the CLEM-LES was found to qualitatively {and quantitatively} capture experimental observations conducted in this study, and also \cite{Kiyanda2013}, in terms of detonation cell structure and irregularity.

Given the success of CLEM-LES for capturing experimental observations of multi-dimensional detonation propagation, the strategy is promising for gaining insight on the physical influence of turbulent mixing rates on other compressible and reactive problems.  For planar detonation propagation in a thin channel, numerical simulations have shown that turbulent mixing rates have significantly influenced the observed cell patterns and detonation structure.  Furthermore, for increased levels of turbulence intensity, such turbulent wave fronts were found to favour local velocities below the CJ-value.  This lead to a favouring of much longer ignition delays, compared to ZND theory, for the unburned gas which passed through the front.  This, in turn, contributed to the formation of unburned pockets in the wake, thus lengthening the overall hydrodynamic structure.  Furthermore, the presence of such unburned pockets was found to give rise to larger and more irregular cell patterns.  These findings, thus, support previous postulates that turbulence gives rise to a large variation of shock-induced ignition delays, and thus the formation of unburned pockets.  In the current work, however, the role of turbulence intensity on the resulting cellular flow patterns, structure thickness, and irregularity were clarified.  All of these qualitative features were found to be significantly altered by the degree of turbulence intensity present.

Finally, in terms of investigation on the role of turbulent mixing itself on highly compressible combustion, it remains an open question why the optimum $C_\kappa$ value to match experiments is higher than theory predicts for incompressible three-dimensional Kolmogorov turbulence.  It is possible that the \hl{predominant} two-dimensional nature of the \hl{simulated} flow field generates more back-scatter, resulting in less subgrid kinetic energy generated at smaller scales.  Furthermore, the compressible nature of the phenomena investigated likely generates back-scatter through amplification of pressure waves that originate from reactive events, which thus feed into larger low-frequency fluid motions \citep{Radulescu2005}.  Thus, compressible problems may inherently require a higher $C_\kappa$ value.  Future numerical investigation should be extended to three-dimensions in order to determine whether a lower $C_\kappa$ value matches the experiments, and to determine how much of an effect compressibility has on $C_\kappa$.\\

Funding for the work presented in this paper has been provided by the Ontario Ministry of Training, Colleges and Universities via an Ontario Graduate Scholarship, a CFD Society of Canada Graduate Scholarship, and NSERC through an Alexander Graham Bell Scholarship, the Discovery Grant, and the H2Can strategic network.  Further acknowledgement is given to Andrzej Pekalski from Shell for financial support in the later stages of the project.  Finally, two academic visits to the University of Leeds, UK, were made possible through additional funding from the University of Ottawa Mobility Scholarship, NSERC Michael Smith Foreign Study Supplement, and also the H2Can strategic network.

\bibliographystyle{jfm}
\bibliography{keylatex}

\end{document}